\documentstyle[12pt,epsfig]{article}

\newcounter{multieqs}

\newcommand{\bq}{\begin{equation}}
\newcommand{\fq}{\end{equation}}
\newcommand{\bqr}{\begin{eqnarray}}
\newcommand{\fqr}{\end{eqnarray}}
\newcommand{\non}{\nonumber \\}
\newcommand{\noi}{\noindent}

\newcommand{\rf}[1]{(\ref{#1})}

\def\zet{\zeta}

\begin{document}

\thispagestyle{empty}
\setcounter{page}{0}


\begin{flushright}
\begin{tabular}{l} 

ANL-HEP-00-0005  \\
hep-th/0001190  \\ 
\end{tabular}
\end{flushright}

\vspace{9mm}
\begin{center}

{\Large \bf S and U-duality Constraints on IIB S-matrices}\\

\vspace{12mm}

Gordon Chalmers \\[5mm]
{\em Argonne National Laboratory \\
High Energy Physics Division \\
9700 South Cass Avenue \\
Argonne, IL  60439-4815 } \\

\vspace{10mm}

{\bf Abstract}
\end{center}

S and U-duality dictate that graviton scattering amplitudes 
in IIB superstring theory be automorphic functions on the 
appropriate fundamental domain which describe the inequivalent 
vacua of (compactified) theories.  A constrained functional form of 
graviton scattering is proposed using Eisenstein series and their 
generalizations compatible with: a) two-loop supergravity, 
b) genus one superstring theory, c) the perturbative coupling 
dependence of the superstring, and d) with the unitarity structure 
of the massless modes.  The form has a perturbative truncation 
in the genus expansion at a given order in the derivative expansion.  
Comparisons between graviton scattering S-matrices  and effective 
actions for the first quantized superstring are made at the quantum 
level. Possible extended finiteness properties of maximally extended 
quantum supergravity theories in different dimensions is implied by the
perturbative truncation of the functional form of graviton scattering 
in IIB superstring theory.

\vfill 
\line(6,0){220} \vskip .04in
e-mail address: chalmers@pcl9.hep.anl.gov \hfill  

\break

\newpage 
\setcounter{footnote}{0}
\section{Introduction}  

S-duality has emerged as a self-equivalence structure of the IIB superstring and 
evidence has accumulated for its existence.  Structures in string theories 
have in the past yielded a better understanding of aspects of 
supersymmetric field theories.  In this work we shall explore 
a manifestly S-dual compatible, perturbative and non-perturbative, expansion 
of the ten-dimensional S-matrix for graviton scattering.   This 
S-matrix is highly constrained by modular forms, and in agreement with 
known results in the low energy limit.  

Recent work on the effective action in the low-energy limit of the superstring 
has revealed the modular property of the S-duality invariant graviton scattering 
to several orders in derivatives \cite{Green:1997tv,Green:1997di}, 
in eleven-dimensional supergravity \cite{Russo:1997mk,Green:1997as, 
Green:1999pu}, and for additional couplings in IIB superstring theory in 
\cite{Berkovits:1998ex}.  We examine the same structure at all orders in 
derivatives and generate these terms through the use of 
modular forms, on the $SL(2,Z)$ fundamental domain for ten and nine-dimensional
theories, and on more complicated domains for lower-dimensional compactified
theories.  In related work there have been $SL(2,Z)$ based reformulations 
at the level of the world-sheet action \cite{Townsend:1997kr,Cederwall:1997ts, 
Chalmers:1999ap} which might lead to a similar description that 
produces higher derivative corrections compatible with S-duality of the IIB 
superstring as for the $R^4$ term \cite{Chalmers:1999ap}.  Tests of S-duality 
at the amplitude level require knowledge of the 
perturbative expansion for graviton scattering at genus greater than one, 
and is formidable.  The S-duality of the superstring imposed on the S-matrix 
leads to non-trivial predictions for its structure, and 
predicts contributions from the genus expansion without performing 
string perturbation theory. 

Standard perturbative expansions involve expansions in the coupling 
constant, or in conjunction with large $N$; however, Feynman diagrams are 
typically difficult at high loop order and alternative approaches are 
useful.   The derivative expansion orderwise must be compatible with the 
non-perturbative symmetry structures 
that exchange weak and strong coupling and must obey these invariance 
properties in IIB superstring theory; possibly this re-ordering avoids  
summation problems, because of factorial dependence in the diagram expansion,   
as it is non-perturbative in the coupling and should thus take into account the 
solitonic contributions.  Such an expansion is 
necessarily non-perturbative from the point of view of the microscopic 
coupling constant in the maximally supersymmetric theory, although perturbative 
in the energy scale for theories without dimensional transmutation.   
M-theory in the eleven dimensional limit requires such an 
approximation, because there is no dilatonic coupling constant in this eleven 
dimensional corner.  

IIB superstring theory has IIB supergravity describing its zero-mode degrees of 
freedom.  Compactification to lower dimensions through tori gives rise 
to several supergravity theories including the example of maximally 
extended $N=8$ supergravity in four dimensions \cite{deWit:1977fk,Cremmer:1978km, 
Cremmer:1978ds,deWit:1982ig}.  The 
remnant of S-duality, and U-duality in general \cite{Hull:1995ys}, imposes severe 
constraints on the perturbative expansions of the supergravity theories in 
various dimensions:  Primarily, we are interested in the finiteness properties 
of the latter theories in this regard.  Recently, $N=8$ supergravity in four 
dimensions has been re-examined, and there is strong evidence that the previously 
thought first primitive divergence of the four-point amplitude does not occur 
at three-loops, but at higher order \cite{Bern:1998ug}.  In specifying the 
supergravity quantum theory from the low-energy limit of the superstring, a 
regulator must be chosen that is compatible with the global symmetries 
of the field equations; on the other hand, superstring theory points to a 
specific regulator, the one in which S-duality remains as a remnant on the 
massless degrees of 
freedom.  At one-loop this is most easily seen in comparing the SUGRA 
expansion with the string and its non-perturbative structure.  

Further divergence nullifications beyond the known properties of  
N=8 supergravity require a field theory mechanism.
Recent work in \cite{Bern:1998ug} has shown that additional cancellations 
not accounted for in superspace powercounting arguments occur.
In the explicit construction of the two-loop amplitudes these 
additional cancellations follow in the cut construction from use 
of on-shell supersymmetry Ward identities.  Alternatively, this 
cancellation occurs because of S-duality, where this structure 
permeates to higher order and in various dimensions.         

In this work we shall reformulate the expansion of the IIB superstring 
graviton scattering using constraints imposed on automorphic functions.  
A uniqueness theorem regarding the functional form incorporating 
additional properties of the scattering elements potentially allows, given 
S-duality, a route to computing complete 
S-matrix elements without perturbative string theory.  A further  
aspect of the modular construction in terms of Eisenstein functions 
implies a truncation property in perturbative supergravity defined 
by the toroidal compactification of IIB in ten dimensions (or $d=11$ 
supergravity).  

This work is organized as follows.  In section 2 we examine the general 
structure of the low-energy limit of the graviton scattering and the 
constraints of S-duality in the uncompactified IIB theory.  In section 
3 we compare the S-matrix with definitions of the quantum effective 
action.  In section 4 we give relevant properties of the Eisenstein 
series used in the construction outlined in section 2.  In section 5 
we analyze toroidal compactifications to lower dimensions and the 
U-duality structure together with the $SL(2,Z)$ subgroup; the emphasis 
in this section is on finding further truncations.  In section 6 
we extract implications for the graviton scattering in the field theory 
limit.  In the last section we give conclusions and discuss extensions 
related to this work. 

\section{S-matrices and Constraints} 

In Einstein frame, S-duality exchanges the coupling constant $\tau=\chi 
+ ie^{-\phi}$ of the IIB superstring with its inverse, and more generally, 
under the $SL(2,Z)$ fractional linear transformation, 
\bq  
\tau\rightarrow {a\tau+b\over c\tau+d} \ .
\fq 
An invariant perturbative expansion for the scattering of gravitons is obtained 
by expanding in the string scale $\alpha'$ at all orders in the 
perturbative series, rather than in the 
string coupling constant $g_s=e^{\phi}$.  Such an expansion is necessarily 
non-perturbative in form from the point of view of the string coupling 
constant, but there are constraints from the known 
structure of the perturbative series in supergravity to be compared with.  

The low-energy expansion of the string S-matrix is found by expanding 
the kinematic invariants parametrizing the scattering at small values 
below the string scale $\alpha'$.  We define the Mandelstam
invariants relevant for the four-point function by 
$s=-(k_1+k_2)^2$, $u=-(k_1+k_3)^2$ and $t=-(k_2+k_3)^2$.  The general 
structure of this  
expansion contains polynomial terms in the kinematic invariants together 
with logarithmic functions, as demanded by unitarity of the massless 
modes.  For $s_{ij}>4/\alpha'$, the unitarity cuts for the massive modes 
of the IIB superstring appear after a  resummation of the former terms; 
an infinite resummation of the higher derivative terms may produce the 
required unitarity cuts for the massive string states.  For example, the 
genus one form for a unitarity threshold of the first massive mode of the 
superstring is $\ln(1-\alpha' s/4)$ and may be expanded at low-energy 
as an infinite series in $s$ via $\sum_{j=1}^\infty (\alpha's/4)^j$.  
The fact that the energy scale of the kinematics is below the first 
massive mode of the string is crucial for preserving the manifest 
S-duality invariant expressions so far found in the literature.  

The string perturbative series is normalized with the conventions  
\bqr  
\kappa_{10}^2={1\over 2} (2\pi)^7 \alpha'^4  
\fqr   
for the ten-dimensional gravitational coupling constant.  Because 
$\alpha'$ enters both in the coupling from the ten-dimensional field 
theory point of view, as well as in the parametrization of the mass 
levels, disentangling of the contributions to the amplitudes from 
the massive modes verse the massless ones of the string needs to be 
carried out.  However, the contributions easily separate to genus 
zero and one in the string perturbative series.  

As of yet, there is no known consistent form of the S-duality compliant 
S-matrix for the IIB superstring in flat ten-dimensional Minkowski background.  
The S-duality invariance of the Einstein frame perturbative expansion demands 
a stringent form of the scattering, for example, of four-graviton scattering 
to low orders in the derivative expansion at fixed, but small, $\alpha'$.  
The work of \cite{Green:1997tv,Green:1997di,Green:1999pv} indicates that 
the form of the scattering of the polynomial terms up to twelve derivatives is 
given non-perturbatively by the series, 
\bqr  
S_{4-point}= \int d^{10}x \sqrt{g}\Bigl[ {1\over \Box^3} R^4 + 
 E_{3/2}(\tau,\bar\tau) R^4 +  E_{5/2}(\tau,\bar\tau) \Box^2 R^4 \Bigr]  \ , 
\label{twotermspoly}
\fqr 
in Einstein frame.  $E_s(\tau,\bar\tau)$ are non-holomorphic Eisenstein series 
(defined in section 4) and the IIB string coupling is $\tau=\chi + i e^{-\phi}$.  
We have included in the first term in \rf{twotermspoly} the massless bosonic 
exchange at tree-level. The string coupling constant is $g_s=e^{\phi}$ and 
the derivatives in the first term is shorthand \rf{twotermspoly} for the 
factor $1/stu$.  

The modular construction of the graviton S-matrix is S-duality invariant, and 
previous attempts to generalize to all orders followed by examination of 
the tree-amplitude for four gravitons 
\cite{Russo:1998fi,Russo:1998vt,GreenMoore}.  However, it gives incorrect 
predictions for the one-loop contribution to the fourteen derivative term 
$\Box^3 R^4$.  

The tree-amplitude for the scattering of four gravitons in a flat background 
in IIB theory is 
\bqr  
A_{4,g=0}^{\rm IIB} = 64 R^4 
  {e^{-2\phi}\over \alpha'^3 stu} {\Gamma(1-{\alpha' s\over 4}) 
 \Gamma(1-{\alpha' t\over 4}) \Gamma(1-{\alpha' u\over 4}) \over 
 \Gamma(1+{\alpha' s\over 4}) \Gamma(1+{\alpha' t\over 4}) 
 \Gamma(1+{\alpha' u\over 4}) } \ , 
\fqr 
or in an alternative form, 
\bqr  
A_{4,g=0}^{\rm IIB} = 64 e^{-2\phi} {R^4\over \alpha'^3 stu} {\rm exp}\Bigl( 
 \sum_{p=1}^\infty {2\zeta(2p+1)\over 2p+1} \left({\alpha'\over 4}\right)^{2p+1} 
 \left( s^{2p+1} + t^{2p+1} + u^{2p+1} \right) \Bigr) \ ,  
\label{treeexpform}  
\fqr 
where the tree-level on-shell effective action \cite{Gross:1986iv} 
found by integrating out massive string modes is contained in the second 
term of the expansion, i.e. $ 2 \zeta(3) e^{-2\phi} \int d^{10}x \sqrt{g} R^4$.
The $R^4$ factor represents the well-known tensor \cite{Green:1982sw}, the 
square of the Bel-Robinson tensor \cite{Bel}, which in linearized $k$-space 
form is found by contracting eight momenta with the external four polarization
vectors, 
\bq  
R^4 = t_8^{\mu_1\ldots \mu_8} t_8^{\nu_1\ldots \nu_8} \prod_{i=1}^4 
\varepsilon_{\mu_i\nu_i} k_{\mu_{i+4}}^i k_{\nu_{i+4}}^i \ ,
\label{supertensor}
\fq 
in momentum space, or in the linear approximation (contributing only to 
the four-point function) of the contraction of four Weyl tensors with the 
tensors $t_8$.  

\begin{figure}
\begin{center}
\epsfig{file=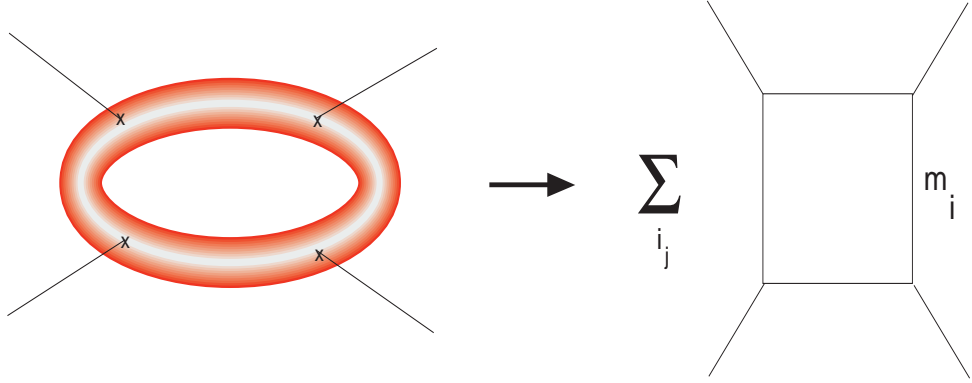,height=5cm,width=13cm} 
\end{center} 
\caption{ 
The low-energy limit of the four-point string genus one 
diagram is represented as an infinite summation of field theory diagrams 
with a string-inspired regulator.} 
\end{figure} 

The conjecture in \cite{Russo:1998vt} for the non-perturbative form 
of the S-matrix is found by replacing the Riemann-zeta functions in 
\rf{treeexpform} with an Eisenstein series through, 
\bq 
\zeta(2p+1) \rightarrow  E_{p+{1/2}}(\tau,\bar\tau) \ , 
\label{zetasub}
\fq 
which takes into account the Einstein frame dependence of the string 
scattering.  Although this 
substitution leads to the correct $R^4$ coupling, it predicts a contribution 
at genus one for the $\Box^3 R^4$ term that is {\em twice} its  
calculated value and is not normalization dependent.  It also does not 
address the required unitarity properties of the massless modes 
(i.e. supergravity).  Building modular invariant functions from 
covariantizing through the substitution in \rf{zetasub} is not unique, 
and we shall modify this conjecture in this work.  

The one-loop amplitude for four-graviton scattering is  
\bqr  
A_{4,g=1}^{\rm IIB} = R^4~ \int \prod_{i=1}^4 d^2z_i~ \int_{{\cal F}_1} 
{d^2\tau\over\tau_2^2}  ~ \prod_{i\neq j} \vert E(z_i,z_j,\tau) 
 \vert^{-\alpha' s_{ij}} e^{-{2\pi\over \tau_2}  k_i\cdot k_j 
{\rm Im}(z_i) {\rm Im}(z_j))} \ ,  
\label{fourpointstring}
\fqr 
with the genus one prime form, 
\bqr  
E(z=z_i-z_j;\tau)= {\Theta\Bigl[{{1\over 2} \atop {1\over 2}}\Bigr] (z,\tau) 
  \over \Theta'\Bigl[{{1\over 2} \atop {1\over 2}}\Bigr] (0,\tau)} \ , 
\label{g1primeform}
\fqr 
and may be similarly expanded in the $\alpha'\rightarrow 0$ limit 
\cite{Green:1999pu}.  
(The denominator in \rf{g1primeform} drops out due to momentum conservation.) 
The low-energy expansion and analytic properties of this amplitude 
is taken in \cite{D'Hoker:1995yr,Chalmers:1998dc,Green:1999pu}, and 
the integral expansion may be directly reproduce as an infinite summation 
of field theory ten-dimensional box diagrams with the appropriate internal 
mass parameters as in figure 4.   In \cite{Green:1999pu} the expansion 
in $\alpha'$ is carried out to order $\Box^2 R^4$ in the derivative (or 
equivalently $\alpha'$) expansion.  We shall need to disentangle the massive 
from the massless mode contributions in this expansion to compare with the 
predicted form of graviton scattering in pure IIB supergravity.  The
massive modes  generate terms in the $q$ expansion of the oscillators when 
the power is non-zero.  It is not that the fundamental 
integration region in \rf{fourpointstring} is demanded by a modular 
cancellation between the massive and massless modes of the superstring;  
modular $SL(2,Z)$ transformations of the punctured torus map regions in 
the field theory limit of the loop integration to other regions in addition 
to those of the massive contributions.  The string, in the field theory 
limit, dictates a well-defined regulator.  In the 
open-string limit suitable for Yang-Mills theory, this complication does not 
enter until the world-sheet corresponds to two-loops.  Although
straightforward at one-loop, technicalities associated with modular  
parameterizations and world-sheet ghosts at higher genus complicate a similar 
expansion at this order and will be analyzed in future work.

We end this section with a clarification of the string-inspired 
regulator at one-loop.  The fundamental domain ${\cal F}_1$ is over 
a restricted domain in the complex plane with a non-trivial region 
near the origin, 

\bqr  
{\cal F}_1 = \Bigl\{ \tau=\tau_1+ i\tau_2 
  : \tau_1^2+\tau_2^2 \geq 1 ~,~ \vert \tau_1\vert \leq {1\over 2} 
 \Bigr\}  \ .  
\label{modregion} 
\fqr 
In field theory the box diagram with masses on the internal lines 
labelled by $j$ is given by the integral representation, 
\bqr  
 {\rm I}_4 (k_j,m_j )&=& \int {d^dl\over (2\pi)^d} ~\prod_j {1\over (l-p_j)^2}   
\non && \hskip -.4in 
 = (2\pi)^{-{d\over 2}} \int \prod_{i=1}^4 da_i~ \delta(1-\sum_{j=1}^4 a_j) 
   \int_0^\infty {d{\tilde\tau}_2\over {\tilde\tau}_2^{-2+d/2}} ~ 
 e^{-{\tilde\tau}_2 [f(a_i;p_i) + a_j m_j^2]}  
\non && \hskip -.4in
=_R (2\pi)^{-{d\over 2}} \Lambda^{3-d/2} \int \prod_{i=1}^4 da_i~  
  \delta(1-\sum_{j=1}^4 a_j) 
 \int_1^\infty {d\tau_2\over \tau_2^{-2+d/2}} ~ 
 e^{-\tau_2 [f(a_i;p_i) + a_j m_j^2]/\Lambda^2}   \ ,
\label{boxdiagram}
\fqr 
where in the last line we have implemented a Schwinger proper time regulator. 
The boundary on the integration is specified through the use of a 
regulator.  In a Schwinger proper time regulator the region for ${\tilde\tau}_2$ 
is be limited to ${\tilde\tau}_2\geq 1/\Lambda^2$.  In dimensional reduction (or 
regularization) we need to integrate over the entire region 
of $\tau_2\geq 0$ with the dimensionally continued integral (in the 
measure).  However, an alernative cutoff scheme for the supergravity 
amplitudes that is consistent with the underlying modular invariance 
of the string is to use the region delimited in \rf{modregion}, 
which is at one-loop found by replacing the integration regime in 
\rf{boxdiagram}, 
\bqr 
\int_1^\infty {d\tau_2\over \tau_2^2} \rightarrow 
 \int_{{\cal F}_1} {d^2\tau\over \tau_2^2} \ .
\label{g1stringinspired}
\fqr 
The region in the complex plane with $\tau_2\geq 1$ is the same, because 
the integration over $\tau_1$ integrates to unity in \rf{boxdiagram}.  
Its difference from the Schwinger proper time cutoff is in a finite region 
in the corner of moduli space.  The integration region in \rf{g1stringinspired} 
should be considered as an alternative regularization scheme, one that is 
compatible in supergravity with the non-perturbative S-duality transformations 
of the superstring.  

The parameterization of the domain of integration in the higher-genus 
moduli space \cite{D'Hoker:1988ta} (and references) 
serves as a string-inspired regulator for higher loop field theory, and gives 
rise to the string-inspired regulated supergravity quantum field theory.  The 
regulated higher genus form is similar to \rf{g1stringinspired} 
but with a more complicated corner region.  The Schwinger proper-time 
regularization generalizes through the additional moduli of the superstring 
as at one-loop.

\subsection{Ten Dimensions}  

The previously conjectured form \cite{Russo:1998vt} for the 
manifestly S-duality invariant function corresponding to the four-graviton 
scattering amplitude used the Eisenstein functions $E_s(\tau,\bar\tau)$ 
to relate the tree-level contributions to the higher genus ones by 
covariantizing the phase in \rf{treeexpform}.  However, it clearly is only 
an approximation for the reason that it predicts the incorrect coefficient 
of the $\Box^3 R^4$ term at genus one, and does not take into account the 
thresholds associated with the massless string modes (i.e. $d=10$ IIB 
supergravity).  In this section, we give a different version that 
incorporates both of these features. 

The simplest modification is to enlarge the set of Eisenstein functions 
to their generalized non-holomorphic and non-modular invariant counterparts.  
The constraints for writing down suitable modular forms and combinations 
follow from : 1) the perturbative series is a power series in even powers of the 
string coupling $\tau_2$, 2) invariance under modular transformations 
in Einstein frame, 3) compatibility with the unitarity structure of the 
massless string modes, 4) the power series predicts the appropriate maximum 
power of $\tau_2$ consistent with the derivative term $\Box^k R^4$ term 
in Einstein frame at tree-level.  We shall also consider only the Eisenstein 
series with half-integer or integer characteristics (the value of which is 
related to the half-integral R-charge shown in \cite{Berkovits:1998ex} and 
\cite{Green:1999by}) and in ten dimensions those 
that converge (lower dimensional compactifications of supergravity have infra-red 
divergences in $d\leq 6$, and possibly the convergence properties are related to 
the infra-red divergences).  The first condition is tight enough to rule out 
most commonly known modular forms, and we shall consider the set of 
generalized Eisenstein functions for $SL(2,Z)$, $E_s^{(q,-q)}(\tau,\bar\tau)$, 
\bqr  
E_s^{(q,-q)}(\tau,\bar\tau) = \sum_{(p,q)\neq (0,0)} 
 {\tau_2^s \over (p+q\tau)^{s+q} 
(p+q{\bar\tau})^{s-q} } \ . 
\label{tendimset}
\fqr 
Their properties are discussed in section 4, together with the 
exceptional modular invariant function 
\bq 
f=\ln \tau_2 \vert\eta(\tau,\bar\tau) \vert^4 
\fq 
which is related to the non-convergent $E_1(\tau,\bar\tau)$ series after 
subtracting the singularity, denoted by ${\hat E}_1(\tau,\bar\tau)$,  
\bqr 
{\hat E}_1(\tau,\bar\tau) = -\pi \ln(\tau_2 \vert\eta(\tau)\vert^4)   \ .
\fqr   
The $SL(2,Z)$ Laplacian acting on $f(\tau,\bar\tau)$ is one,  
\bq 
\tau_2^2 \partial_\tau \partial_{\bar \tau} \ln \tau_2 \vert\eta(\tau,\bar\tau) 
\vert^4 = 1 \ , 
\label{excepmod}
\fq 
which in the string setting represents contributions from only one 
perturbative order (the form is required for genus one and for the 
$\Box \ln \Box ~R^4$ tensor).  Our ansatz 
consists of polynomials in this combined set.  The cusp forms are modular 
invariant functions that vanish at $\tau_2\rightarrow \infty$ and 
admit the expansion, 
\bqr  
f_{\rm cusp}(\tau,\bar\tau) = \sum_{n\neq 0} a_n \tau_2^{1\over 2} K_{n-1/2} 
(2\pi \vert n\vert \tau_2) e^{2\pi i n\tau_1} \ ,
\fqr 
expressed in terms of exponentials of $\tau_2$.  They do not contribute to 
the perturbative expansion of either the superstring or supergravity, and 
we shall not consider them or the modifications of the polynomial terms by 
them in any detail.  The first instanton correction 
to the $R^4$ term in \cite{Green:1997tv} provides evidence that they 
do not contribute to this order, and a similar calculation at higher 
$\alpha'$ is necessary to predict whether or not they contribute to the 
higher derivative terms.  Explicit forms for the cusp forms on the fundamental 
domain $U(1)\backslash SL(2,R)/SL(2,Z)$ are not known \cite{Terras}.  Similarly, 
the ones relevant for the moduli spaces of the toroidally compactified theories 
are not known (these spaces contain the former as a subspace).  

We first discuss the polynomial terms arising in the low-energy expansion, 
followed by the non-analytic (logarithmic) ones.  
The transformation from string frame to Einstein frame, $\eta_{\mu\nu}^{(s)} 
= \tau_2^{1\over 2} \eta_{\mu\nu}^{(E)}$, induces a coupling constant 
into the derivative expansion that is important for the $\tau_2$ counting.  
For example, the $\Box^k R^4$ terms in the string frame at tree-level 
are described in Einstein frame, 
\bq 
A_{4,g=0}^{IIB, k} = N_{g=0}^k
\zeta({3\over 2}+{k\over 2}) \tau_2^{3/2+k/2} \Box^k_{(E)} R^4_E  \ ,  
\fq  
for $k\neq 1$, and 
where the $e^{-2\phi}$ string coupling inherit in \rf{treeexpform} and 
the transformation of the tensors are taken into account.  
This power of $\tau_2$ is the maximum one possible in the perturbative 
series and successive higher genus contributions lower it by two units 
successively.  The general form of the contribution at arbitrary $k$ 
to all genus,  
\bq  
A_4^{IIB,k} = f_k(\tau,\bar\tau) ~\Box^k_{(E)} R^4_{(E)} \ , 
\fq 
demands the structure of $f_k(\tau,\bar\tau)$ to have the form, 
\bq  
f_k(\tau,\bar\tau) = \zeta({3+k\over 2}) \tau_2^{{3\over 2}+{k\over 2}} 
+ a_1 \tau_2^{-{1\over 2}+{k\over 2}} + a_2 \tau_2^{-{3\over 2}+{k\over 2}} 
+ \quad \ldots \quad + {\cal O}(e^{2\pi i \tau}) \ .  
\label{kform}
\fq 
The coefficients $a_g$ are constants that may be found by doing explicit 
low-energy string perturbation theory calculations up to genus $g$.  We will 
see that given the set of modular forms discussed above the series $a_k$ 
receives non-vanishing values up to a maximum genus for a given $k$.  

The fact that only one tensor, namely the eight-derivative $R^4$ covariantized
from the linear form in \rf{supertensor}, appears in the description of the
four-point  function is consistent both with perturbative IIB superstring theory
up  to genus four \cite{Lechtenfeld:1990ke} as well as with constraints in 
supergravity from unitarity constructions.  This property is due to
$N=8$ supersymmetry at one \cite{Green:1982sw} and two-loops and also to higher
orders \cite{Bern:1998ug}: The uniqueness of this tensor structure is the
primary influence of $N=8$ supersymmetry in constraining the S-matrix as the loop
integrations are not constrained directly by supersymmetry without additional
regulator specification (such as choosing one compatible with the global
duality symmetries of the classical field equations of maximal supergravity).  
Lower derivative tensor structures appear in theories with lower supersymmetric 
gravitational theories \cite{Harvey:1998ir}.  

For $k=0$, the only function in the set considered that could describe 
the expansion is $E_{3/2}^{(0,0)}$ (modulo cusp forms); the $R^4$ term then 
receives perturbative contributions at tree-level and one-loop.  Explicit 
calculations in supergravity indicate that, at two-loops and higher, an 
additional $\Box^2$ may explicitly be pulled out from within the loop
integrations  (in figure 2 we list the integral contributions, where 
the tensor $R^4$ is not displayed).   
Thus, this truncation is consistent with known results (at tree and 
one-loop the massive modes in the $q$ expansion explicitly are of order 
$\alpha'$, hence $\Box$ by dimensional analysis, higher than the massless 
modes).  The value 
$k=1$ is special because momentum conservation of the four external 
gravitons, through $s+t+u=0$, forces the on-shell ten-derivative polynomial 
term to vanish.  At $k=2$, again there is only one function one may write down, 
namely $E_{5/2}(\tau,\bar\tau)$, and predicts a genus zero and genus 
two contribution but none from genus one; explicit 
string one-loop calculations shows the vanishing of the genus 
one coefficient.  
 
\begin{figure}
\begin{center}
\epsfig{file=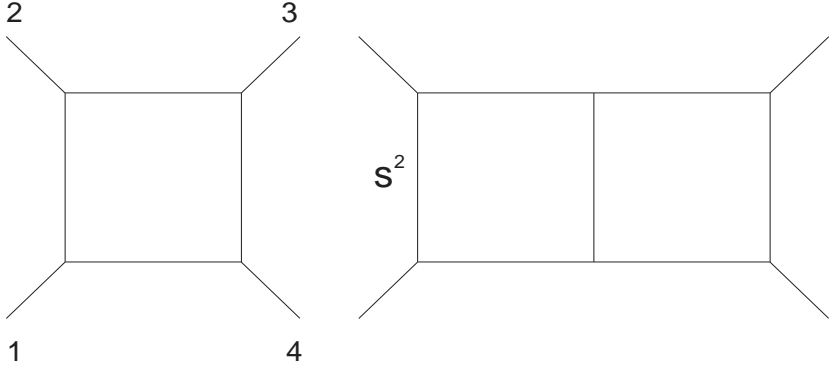,height=5cm,width=11cm} 
\end{center} 
\caption{ 
Sample supergravity field theory contributions to the one- and 
two-loop graviton scattering before symmetrization.  To these 
orders the derivatives are explicitly extracted from the tensor 
integrations.}  
\end{figure}  

At values of $k\geq 3$ a break with the previous pattern emerges  
because the non-modular invariant $E_s^{(q,-q)}(\tau,\bar\tau)$ 
for $q\neq 0$ may contribute.  At $k=3$ we may introduce the form 
\bqr 
f_3(\tau,\bar\tau) = \alpha_1~ E_{3/2}^2(\tau,\bar\tau) + \alpha_2 ~ 
E_{3/2}^{(1,-1)}(\tau,\bar\tau) E_{3/2}^{(-1,1)}(\tau,\bar\tau) \ ,   
\fqr 
(where $E_{3/2}^{(-1,1)}=E_{3/2}^{\star(1,-1)}$).  It has the 
large $\tau_2$ perturbative expansion, 
\bqr  
f_3(\tau,\bar\tau)=4 (\alpha_1+\alpha_2)\Bigl[ \zeta(3)^2 \tau_2^3 +  
\left({\alpha_1-{\alpha_2/ 3}\over \alpha_1+\alpha_2}\right) 4\zeta(2)\zeta(3) \tau_2 
+ \left({\alpha_1-{\alpha_2/ 9}\over \alpha_1+\alpha_2}\right) 4 \zeta^2(2) 
\tau_2^{-1} \Bigr] \ . 
\label{f3function}
\fqr
The relative coefficient between $\alpha_1$ and $\alpha_2$ is chosen to agree 
with the genus one contribution for $\Box^3 R^4$.  
Note that \rf{f3function} differs from previous forms because of the introduction 
of the modular invariant contribution of the generalized Eisenstein functions.  
A relative value of $\alpha_1={5\over 3}\alpha_2$ leads to agreement with the 
calculated genus zero and one coefficients in string theory; this combination 
predicts a non-vanishing coefficient at genus two and {\em no further contributions}.  

\begin{table} 
\caption{ 
The marks indicate which modular forms could contribute 
at genus $g$ to the $\Box^4 R^4$ term in the four graviton scattering 
amplitude.}  
\begin{center}
\begin{tabular}{|l|l|l|l|l|r|} \hline 
{\em }     &  $g=0$ & $g=1$ & $g=2$ & $g=3$ \\ \hline 
$\alpha_1$ & $\surd$ &  &  & $\surd$    \\ \hline    
$\alpha_2$ &  & $\surd$ & $\surd$ &   \\ \hline 
\end{tabular} 
\end{center} 
\end{table}

Combinations of these modular forms is straightforward to construct 
at higher derivatives, although it is not uniquely determined through direct 
covariantization via the substitution in \rf{zetasub}.  Futhermore, in contrast 
to the $k\leq 3$ cases, such combinations allow for perturbative contributions 
arising from modular forms that contribute at orders of genus solely 
above tree-level.  The non-analytic terms in the low-energy expansion 
require such combinations.  

\begin{table} 
\caption{Example contributions to $\Box^k R^4$ arising in string 
perturbation theory at genus $g$ from the structure of the 
generalized Eisenstein series.} 
\begin{center}
\begin{tabular}{|l|l|l|l|l|l|r|} \hline 
{\em } &  $g=0$ & $g=1$ & $g=2$ & $g=3$ & $g=4$ & $\ldots$ \\ \hline 
$R^4$      & $\surd$ & $\surd$ & & & & \\ \hline  
$\Box R^4$ &         &         & & & & \\ \hline  
$\Box^2 R^4$ & $\surd$ &   & $\surd$ & & &  \\ \hline 
$\Box^3 R^4$ & $\surd$ & $\surd$ & $\surd$ & & & \\ \hline 
$\Box^4 R^4$ & $\surd$ & $\surd$ & $\surd$ & $\surd$ & & \\ \hline 
$\Box^5 R^4$ & $\surd$ & $\surd$ & $\surd$ & $\surd$ & & \\ \hline 
$\cdots$ & & & & & & \\ \hline     
\end{tabular} 
\end{center} 
\end{table}

As another example, consider $k=4$ and $k=5$.  In this case we may use, 
\bqr  
f_4(\tau,\bar\tau) = \alpha_1~ E_{7/2}(\tau,\bar\tau)  
 + \alpha_2 ~ E_{3/2}(\tau,\bar\tau) \ , 
\label{f4modular}
\fqr 
and,  
\bqr  
f_5(\tau,\bar\tau) &=&  \beta_1~ 
E_2(\tau,\bar\tau) E_2(\tau,\bar\tau) + \beta_2~ E_2(\tau,\bar\tau)    
+ \beta_3~ E_{3/2}(\tau,\bar\tau) E_{5/2}(\tau,\bar\tau) 
\non && 
+ \beta_4~ E_2^{(1,-1)}(\tau,\bar\tau) E_2^{(-1,1)}(\tau,\bar\tau)  
+ \beta_5 E_{3/2}^{(1,-1)}(\tau,\bar\tau)  E_{5/2}^{(-1,1)}(\tau,\bar\tau) 
+ {\rm c.c.} \ . 
\label{f5modular}
\fqr 
There is one condition on coefficients $\beta_1$, $\beta_2$, $\beta_4$ in 
the asymptotic expansion from the perturbative expansion, that the 
$\tau_2^{-1}$ term is zero. 
The number of different combinations increases with larger $k$ values.   
Similar functions may be constructed to all orders, and the functional 
form at higher derivatives is one of the central results in this work.  
In \rf{f4modular} the different combinations generate perturbative
corrections at the following genus orders tabulated in Table 1. 

The perturbative series of the polynomial terms for the $\Box^k R^4$ 
tensor in string theory, using the generalized Eisenstein functions 
in \rf{tendimset} for $s=n/2$ and $n$ integer in a polynomial fashion, 
receives corrections 
up to maximum  $g_{\rm max}={1\over 2}(k+2)$ genus in string theory 
for $k$ even and $g_{\rm max}={1\over 2}(k+1)$ genus for $k$ odd,  
listed in Table 2.  Higher 
genus calculations in IIB superstring theory is necessary to verify 
the form as well as to make agreement with the coefficients; an 
indirect resummation of the leading terms in $\tau_2$ of the derivatives 
must also agree with the thresholds of the massive modes of the 
superstring and might fix the coefficients.  

In the remainder of this section we examine other functions 
built out of Eisenstein series.  Generic combinations of the 
generalized Eisenstein series  
either do not give rise to the appropriate powers of $\tau_2$ in 
accord with the string perturbative series or have values of 
$s\leq 1$ and not converge.  Ratios of Eisenstein series in the 
large $\tau_2$ perturbative regime, for example, 
\bqr  
{E_{3}(\tau,\bar\tau)\over E_{3\over 2}(\tau,\bar\tau)} &=& 
 { {a_{3} \tau_2^{3} + 
 b_{3} \tau_2^{-2} + n_1 e^{-2\pi \tau_2} + \ldots} \over  
 { {a_{3/2} \tau_2^{3\over 2} + b_{3/2} \tau_2^{-{1\over 2}} + m_1  
  e^{-2\pi\tau_2} + \ldots } } } 
\ ,
\label{ratioone} 
\fqr 
either do not generically reflect the appropriate $\tau_2$ dependence of string 
perturbation theory, or else,   
\bqr  
{E_{3/2}^3(\tau,\bar\tau)\over E_{5/2}(\tau,\bar\tau)} = 
{(a_{3/2} \tau_2^{3 \over 2} + b_{3/2} \tau_2^{-{1\over 2}} + e^{2\pi i\tau} 
+ \ldots )^3 \over a_{5/2} \tau_2^{5\over 2} + b_{5/2} \tau_2^{-{3\over 2}} 
+ c e^{2\pi i\tau} +\ldots} 
\label{ratiotwo}
\fqr 
generate at large $\tau_2$ perturbative $\tau_2$ dependence in accord with 
string perturbation theory and could potentially be used to describe the 
$\Box^5 R^4$ term.  However, the dependence of $\tau_2$ on the exponential 
terms generically does not match with expectations of the D-instanton series 
(which begin with a single order one factor for the first correction 
after expanding \rf{ratiotwo}).  The function in \rf{ratiotwo} has dependence 
on half-integral powers of $\tau_2$, and its exponentials are suppressed by 
a power of $\tau_2^{5/2}$.  Rigorously, we can not rule out the ratio 
combinations similar to that in \rf{ratiotwo} at all orders without further
constraints.   

Non-half integral or integral values of $s$ also generically lead to 
dependence on $\tau_2$ which is not captured by perturbation theory.  
For example, the product 
\bqr 
E_{5/4}^2(\tau,\bar\tau) &=& \bigl(a\tau_2^{5\over 4}+ b \tau_2^{-{1\over 4}}+ 
{\cal O}(e^{2\pi i\tau}) \bigr)^2  
\non && 
= a^2 \tau_2^{5\over 2} + 2ab \tau_2 + b^2 \tau_2^{-{1\over 2}} +  
 {\cal O}(e^{2\pi i\tau})  \ , 
\fqr 
has a $\tau_2$ factor which is not suppressed by a factor of $\tau_2^2$ 
relative to the first term and cannot arise in string perturbation theory.  
Generic products of Eisenstein series with non-half integral values of $s$ 
also have $\tau_2$ dependence that does not agree with the perturbative series.      

Any of the functions built out of the products of generalized Eisenstein 
series may also be used to generate further
automorphic functions  by acting on them with the covariantized Laplacian
$\nabla=\tau^2_2 \partial_\tau \partial_{\bar\tau}$, e.g. $\nabla
E_{3/2}^2(\tau,\bar\tau)$.  The polynomial action of $\nabla$ generates the
same perturbative truncation in the modular ansatz up to a maximum genus 
for a given $\alpha'$ order, although the instanton corrections may be 
modified together with cusp forms.  Their effect on the perturbative 
series is to adjust coefficients in the perturbative series up to the 
maximum genus.  The action up to $\Box^3$ order in the 
derivative expansion does not introduce additional automorphic functions 
because the single Eisenstein functions are eigenfunctions of $\nabla$, 
and their use at higher derivatives appears redundant perturbatively. 

In lower dimensional toroidally compactified theories, the limiting 
Eisenstein functions relevant for describing the $\Box^k R^4$ terms 
have less well behaved convergence and are possibly 
related to the infra-red divergences in $d\leq 6$ supergravity.  The general 
structure of the perturbative series arising through the modular form 
construction required by S-duality invariance is listed in table 2.   

The pre-factors $E_{3/2}(\tau,\bar\tau)$ and $E_{5/2}(\tau,\bar\tau)$ of 
the low-energy polynomial expansion up to $\Box^2 R^4$ satisfy Laplacian 
eigenvalue conditions.  We are not imposing a property  
\bqr  
4 \tau_2^2 {\partial\over\partial\tau} {\partial\over \partial{\bar\tau}} 
f^{(i)}_k(\tau,\bar\tau) = \lambda^{(i)}_k f_k^{(i)}(\tau,\bar\tau) \ , 
\label{laplacian}
\fqr 
on terms $f^{(i)}_k$ of $f_k(\tau,\bar\tau)$ or a similar one on 
the potentially covariantized phase in \rf{treeexpform}, which 
together with the asymptotic behavior in \rf{kform}, limits 
only single Eisenstein series and their generalizations to being a 
solution (together with cusp forms) to \rf{laplacian} \cite{Terras}.  
The $SL(2,Z)$ covariantization of the leading asymptotic term in $f_k$ 
does satisfy a Laplacian condition, i.e. $\nabla (\tau_2^{{3\over 2}+{k\over 2}}) 
= 4({3+k\over 2})({1+k\over 2}) (\tau_2^{{3\over 2}+{k\over 2}})$, but 
may be covariantized with $SL(2,Z)$ in different ways.  For example, 
the simplest gives $E_{{3\over 2}+{k\over 2}}(\tau,\bar\tau)$ but also 
$E^2_{{3\over 4}+{k\over 4}}(\tau,\bar\tau)$ or 
$\vert E^{(1,-1)}_{{3\over 4}+{k\over 4}}(\tau,\bar\tau) \vert^2$.  A 
naive covariantization of the leading terms is not unique and for 
this reason the graviton scattering then is not going to exponentiate 
directly into the form of a tree-like amplitude.  

\subsection{Non-analytic terms}

The unitarity structure of the perturbative superstring amplitudes also 
requires a description in terms of modular forms in order to construct 
manifestly S-dual scattering.  Clearly such a construction is different 
than the above because at tree-level there are no non-analytic contributions 
and the perturbative expansion involving logarithmic functions, for example 
$\ln^L(\Box)$ and $\ln\ln\ldots\ln\Box$ (up to L iteratively), begins 
at genus one.  

The form of the action containing higher orders in derivatives 
depends on the choice of the definition of the effective action 
but is uniquely defined by an S-matrix compuation in string theory.  
As required by unitarity, the S-matrix and the effective action 
will both contain logarithmic terms.  At one-loop, for example, 
explicit expressions found from expanding the integrated four-point 
S-matrix element has a contribution of the form 

\bqr 
C_{\rm log}^{g=1} = \Bigl( s \ln s + t \ln t +u \ln u \Bigr) R^4 .  
\label{examplelog}
\fqr 
Higher logarithms appear at multi-loop and similar structure is required 
in higher-point gravitational amplitudes.  Under the transformation to 
Einstein frame from string frame the kinematic invariants pick up factors 
$s_s \rightarrow s_e \sqrt{\tau_2}$.  (The metric is rescaled by a square-root 
factor of the coupling constant in ten dimensions.)  The logs potentially produce 
further non-analyticity in the string coupling constant,  

\bqr 
s\ln s \rightarrow \sqrt{\tau_2} s_e \ln ( \sqrt{\tau_2} s_e ) . 
\fqr 
In Einstein frame, the coupling $e^{-\phi}=\tau_2$ does not 
just count loops via the the topogical coupling, 
\bqr  
S_{\rm dil}=\int d^2z \sqrt{g} R \langle \phi\rangle = 2(1-g) \phi \ ,  
\fqr 
but rather includes a dilatonic factor within the free superstring 
action.  S-duality does not simply exchange weak coupling with strong 
coupling in the Einstein frame for individual modes of the string:  In 
a field theory setting the the coupling constant is inverted, but 
also the contributions of the massive modes of the superstring get 
mixed differently than the massless ones within the integration because 
of the scale factor introduced in transforming to Einstein frame.  
The $\tau_2$ dependence in the massive modes is not removed in the 
space-time propagator $\sqrt{\tau_2} s_{e} - m_i^2$.    

In \rf{examplelog} because of momentum conservation, the $\tau_2$ 
dependence in the logarithm cancels out $(s+t+u=0)$.  The unitarity 
construction below is independent of $\tau_2$ dependence in the 
non-analytic (i.e. logarithmic) terms at higher loops, which iteratively 
constructs $SL(2,Z)$ invariant non-analytic terms.  

The on-shell polynomial terms in previous sections are sufficient to determine, 
through supersymmetrizing the {\it on-shell} $\Box^k R^4$ tensor, these 
non-analytic terms through a unitarity construction.  The imaginary part 
in a particular channel, for example $s=-(k_1+k_2)^2$, of the four-graviton 
scattering amplitude is determined through the unitarity relation 

\bqr  
{\rm Im}_s~ A_4 (k_i) = \sum_{n=2}^\infty \sum_{\lambda_j} &\int& d\phi_n ~ 
 A_{n,\lambda_j}\left[ g(k_1),g(k_2); p({\tilde k}_1), p({\tilde k}_2), 
 \ldots, p({\tilde k}_n)\right]  \non && \times 
 A_{n,\lambda_j}^\star \left[ g(k_3),g(k_4); p({\tilde k}_1), p({\tilde k}_4), 
  \ldots, p({\tilde k}_n) \right] \ ,
\label{cutconstruct}  
\fqr 
with the $n$-body phase space integration measure given by, 
\bqr 
d\phi_n =\prod_{j=1}^n {d^d{\tilde k}_j\over (2\pi)^d}  
  \delta(\sum_{i=1}^n {\tilde k}_i+\sum_{j=1}^4 k_j ) 
  \prod_{j=1}^n \delta^{(d)}({\tilde k}_j^2) \Theta({\tilde k}_j^0)  \ ,
\fqr 
and where $\lambda_j$ denotes the quantum numbers of the physical states ``$p$'' 
of the intermediate lines (gravitons and their supersymmetric partners in the 
gravitational multiplet).  The phase space integration is over the region 
\bqr  
{\tilde k}_j^2=0 \qquad \sum_{i=1}^4 k_i+\sum_{j=1}^n {\tilde k}_j=0  \ , 
\fqr 
of out-going momenta.  The equation in \rf{cutconstruct}  
gives an iterative construction of the non-analytic terms in the 
four-graviton scattering amplitude.  At energies $s\leq 4/\alpha'$ only 
the massless modes of the superstring contribute to the imaginary part in 
the s-channel.  Furthermore, the {\it on-shell} supersymmetrization of the 
$\Box^k R^4$ terms (together with a similar construction of higher-point graviton 
scattering amplitudes) provides the necessary amplitudes $A_n$ that are 
to be inserted into \rf{cutconstruct}; a on-shell linearized IIB superspace 
is known \cite{Howe:1984sr} in the absence of an off-shell one which might 
aid in the explicit supersymmetrization to higher order.  Given an 
S-duality compliant form of the higher derivative polynomial terms, the 
non-analytic ones are necessarily also invariant under $SL(2,Z)$ transformations 
through \rf{cutconstruct}.  Higher-point graviton scattering amplitudes may also 
be S-duality covariantized which is an ingredient in \rf{cutconstruct}.  

We may expand the amplitudes in \rf{cutconstruct} to a total order $m$ 
which receives contributions from the respective expansions of the two 
amplitudes to all orders $k$ and $l$ so that $k+l=m$.  The order $\alpha'^m$
non-analytic term is then determined from the polynomial ones together 
with the lower-order non-analytic ones.  In this iterative manner, the 
non-analytic terms in the low-energy expansion of the gravitational 
S-matrix is $SL(2,Z)$ invariant if the polynomial terms are.  

\section{Effective action and S-matrix in derivative expansion}

We shall re-examine the eight-derivative $R^4$ term in this section 
to review its form in view of the S-matrix; the off-shell effective 
action for the IIB superstring is difficult to define because off-shell 
string scattering and the action for the five-form self-dual field strength 
off-shell are not available directly although 
several different on-shell effective action constructions may be given.  
In this section we examine primarily the four-point function, the form 
of which does not alter significantly at one-loop; however, at higher-point 
because of contributing independent high-point diagrams from boxes to 
$n$-gons ($n\leq d$) there will be significant differences at genus one and 
higher.  

The on-shell effective action presented in the literature up to eight 
derivatives \cite{Gross:1986iv,Green:1997tv} is virtually indistinguishable from 
the S-matrix due to supersymmetry.  
There are two differences between different functional form of an on-shell 
effective action to this derivative order.  First, the term 
arising from massless string exchange at tree-level, i.e. in $\alpha' 
\rightarrow 0$ limit the graph in figure 3 with intermediate boson lines, 
is not included.  Second, there is regulator dependence in the quantum 
one-loop contribution in the field theory setting (entirely from the massless 
modes in a $q$-expansion) that breaks the S-duality invariance 
if a string-inspired regulator is not used in comparing the supergravity 
with the string.   

Furthermore, the linearized $N=8$ (non-linear) supersymmetry will cancel 
any triangle or bubble sub-graph in a multi-loop graviton scattering 
amplitude.  The string diagrams then, in the field theory limit, that 
have been included in the definition of the IIB effective action are 
identical to the ones that contribute to the four-point scattering 
amplitude in IIB superstring theory.  Although at the 
four-point level the only difference between the S-matrix and the 
effective action in \cite{Green:1997tv} is contained in 
\rf{masslesstree}, at higher-point there are contributions 
of external massless trees to non-vanishing loop diagrams and then 
more differences associated with the external massless trees from 
a definition adopted in \cite{Sakai:1987bi}.  Agreement 
with results in \cite{Berkovits:1998ex} is obtained in an S-matrix calculation 
in superstring theory.  
 
\begin{figure}
\begin{center}
\epsfig{file=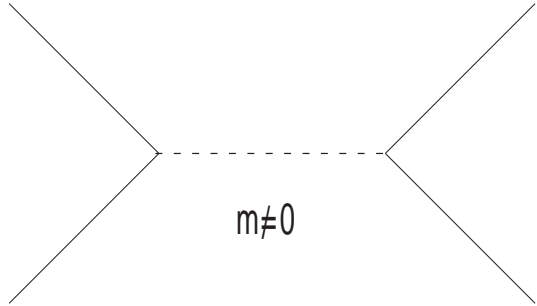,height=4cm,width=7cm} 
\end{center} 
\caption{ 
The diagrams with intermediate massive string 
modes contributing to the tree-level graviton exchange.} 
\end{figure} 

In this remainder we examine the differences of the different 
definitions together with S-duality at the eight derivative order.  
In Einstein frame, the massless tree-exchange, identical to that 
in figure 3 but with intermediate bosonic modes together with the 
four-point vertex, gives rise to a covariantized contribution, 
\bq 
A= {1\over \kappa_{10}^2} \int d^{10}x \sqrt{g}~ {1\over\Box^3} R^4 \ ,  
\label{masslesstree}
\fq 
where the $\Box^3$ represents the appropriate combination of derivatives 
to produce a $1/stu$ in momentum space, and is independently invariant 
under the S-duality transformation.  In not keeping it, the 
effective action is thought of as giving rise to 1PI quantum corrected 
vertices which must be sewn together to generate an S-matrix, i.e. 
illustrated as shaded circles in figure 3, and requires an off-shell 
generalization to construct the S-matrix.  Furthermore, at one-loop 
there is an additional $SL(2,Z)$ of modular invariance and in not 
keeping the term in \rf{masslesstree} arising from the massless sector 
breaks this in the string scattering perturbation theory.  
To this order the terms that break modular invariance do not break with 
S-duality, however, at higher-order this is possible.   

We examine a more precise definition for the S-duality invariant graviton 
scattering expression in the following - different definitions 
of the low-energy quantum effective action will produce further differences 
at the four-point level at order eight derivatives than just that in 
\rf{masslesstree} and potentially break the S-duality.  The S-duality 
invariant expressions that have been obtained so far demand that the 
external kinematics are below the string scale, e.g. $s_{ij}\leq 4/\alpha'$ 
despite the fact that the massless and massive modes are being integrated 
out at the quantum level.  Not keeping the term in \rf{masslesstree} gives 
a definition similar to a combination of a one-particle irreducible for the 
massless quantum fields together with one-particle reducible for the massive 
modes.  In examining the implications of S- and U-duality on the supergravity 
theory the regulator must be chosen in a way that is compatible with these 
symmetries of the classical field equations.  

Supersymmetry together with M-theory constraints suggest that the coefficient 
of the $R^4$ term is a function $f$ satisfying a Laplacian condition on the 
fundamental domain of $U(1)\backslash SL(2,R)/SL(2,Z)$: $\tau_2^2 \partial_\tau 
\partial_{\bar\tau} f= 3/4 f$.  Boundary conditions coming from the perturbative 
one-loop calculation need to be specified in order to find the solution.  In 
supergravity the domain of vacua is the entire complex $\tau$-plane of couplings 
as S-duality is not a structure only in the massless sector of the string only.  

The general form of the tree- and one-loop contributions to the $R^4$ term are 
of the type, 
\bqr  
f_{\rm pert}(\tau,\bar\tau) = a\tau_2^{3/2} + b \tau_2^{-1/2} \ ,
\fqr 
and arise from massive mode exchange at tree-level and massless ones at 
one-loop (with coefficients $a$ and $b$ respectively).  The $SL(2,Z)$ 
invariant completion of the former term gives a form, 
\bqr 
f(\tau,\bar\tau) = a\tau_2^{-1/2} + b \sum_{p,q}  
  {\tau_2^{3/2}\over \vert p+q\tau\vert^3} 
\label{generalf}
\fqr 
and agrees with perturbative IIB superstring theory when, 
\bqr 
f=2\zeta(3) \tau^{3/2}_2 + 4\zet(2) \tau_2^{-1/2} + {\cal O}(e^{2\pi i\tau}) \ ,  
\label{ffunction} 
\fqr 
where the exponentially suppressed terms correspond to $k$-multiple D-instanton 
corrections in the superstring theory (for example, calculated for $k=1$ in 
\cite{Green:1997tv}).  The effective actions found by distinguishing the 
massless modes of the superstring with different treatments changes the first 
coefficient $a$ in \rf{generalf} to different values.  
It measures the regulator influence in the supergravity, that is, the 
cutoff $\Lambda$ dependence or the use of dimensional regularization.  Only for 
one value is the result in \rf{generalf} $SL(2,Z)$ invariant, $a=0$, and that 
is the one that comes directly from either an S-matrix element calculation in 
IIB string theory or through supergravity with a regulator modelling the 
one that arises in the low-energy limit of the quantum superstring.   

In supergravity at one-loop the massless modes contribute, 
\bqr 
I(s,t,u)= \int {d^dl\over (2\pi)^d} ~{1\over 
 l^2 (l-k_1)^2 (l-k_1-k_2)^2 (l+k_4)^2 } + u\leftrightarrow t + 
 s\leftrightarrow u  \ , 
\label{oneloopgrav}
\fqr 
to the $R^4$ tensor and must be regulated because it is quadratically divergent 
in ten dimensions.  In dimensional reduction ($d=10-\epsilon$) the result is,  
\bqr 
I(s,t,u) =  C_\epsilon \Bigl[ {1\over \epsilon} (s+t) +  {1\over \epsilon} (u+s) + 
 {1\over \epsilon} (t+u) \Bigr] + ({\rm non-analytic}) \ ,  
\label{dimred} 
\fqr 
and gives no contribution to $R^4$; dimensional reduction breaks S-duality.  
However, in a string-inspired regulator \rf{g1stringinspired} the result from 
\rf{oneloopgrav} is  
\bqr  
A_4^{[N=8]}(k_i,\epsilon_i) \sim  R^4 \Bigl[ \Lambda^2 + (s^2 +t^2+u^2) 
 \ln \Lambda^2 + ({\rm finite}) \Bigr] \ .
\label{stringfieldoneloop}
\fqr 
The quadratic divergence in a string-inspired regulator is roughly proportional 
to the inherit string scale $\Lambda^2 \sim 1/\alpha'$.  The precise coefficient 
of the one-loop contribution to the $R^4$ term in $d$-dimensions 
depends on the regulator chosen and agrees with the string result for the 
$R^4$ term when the domain in \rf{g1stringinspired} is taken.  Scaling 
the coupling constants to force agreement between a general integration 
in \rf{stringfieldoneloop} changes the instanton corrections predicted 
from the factor $E_{3/2}$ of the $R^4$ term.  The calculation may 
be straightforwardly generalized to arbitrary dimensions and to its ultra-violet 
finite loop integration when $d<8$.  

The string-inspired regulator from the field theory point 
of view makes the calculated result from \rf{stringfieldoneloop} agree with 
S-duality.  Changing the regulator with the use of dimensional reduction 
gives a different value, $c=0$ in \rf{ffunction} from \rf{dimred}.  The 
above regulators preserve (non-linear) supersymmetry to this 
order, but only one gives rise to an explicit $SL(2,Z)$ invariant expression, 
namely the string-inspired regulator built into the supergravity theory.  

S-duality is not expected to be a property of the low-energy IIB supergravity 
theory, as may be explicitly found to the order eight derivative $R^4$ term by 
dropping the massive mode contributions which contribute the $\tau_2^{3/2}$ 
term in \rf{generalf}.  However, as a structure in the IIB superstring, it 
survives as a remnant in the IIB supergravity theory directly if a 
regulator is chosen so that the integration region in a first quantized form 
is chosen to mimic that of the superstring.  At one-loop this corresponds to 
using a Schwinger proper time form of the four-point function in \rf{g1stringinspired}.   
  
There is a one parameter family of functions $f_{\rm pert}(\tau,\bar\tau)$ 
consistent with supersymmetry that leads to the form in \rf{generalf},  
and the definition that arises from the low-energy limit of 
the string amplitude or in supegravity with the regulator chosen in 
\rf{g1stringinspired} agrees with the S-duality invariant form when 
$a=0$ in \rf{generalf}.  At two-loops the supegravity modular parameterization 
and regulator has 
been elucidated in \cite{Green:1999pu}.  Explicit calculations of 
one-loop four-graviton scattering in IIB supergravity indicate that all 
the different definitions of the effective action are encoded in one 
term to order eight derivatives.  

In deducing results regarding the perturbative series of maximal supergravity 
directly from the superstring a multi-loop regulator must be chosen to 
agree with the modular properties of the IIB superstring S-matrix.  
 
\begin{figure} 
\begin{center}
\epsfig{file=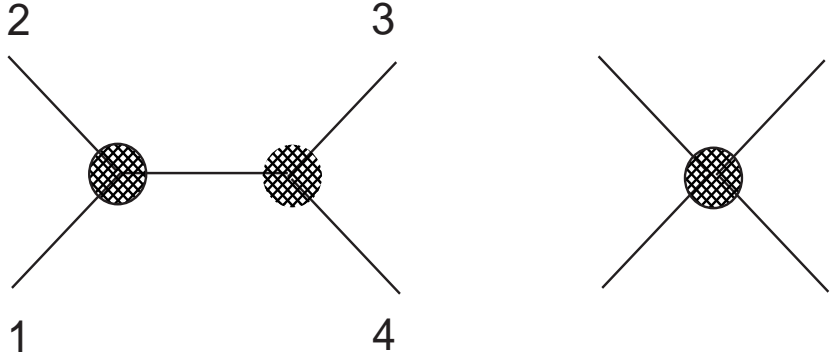,height=5cm,width=11cm} 
\end{center} 
\caption{ 
The circles denote quantum corrected effective 
vertices derived from an effective action.  Modular invariance 
is regained after sewing to obtain trees.} 
\end{figure} 

\section{Modular Forms: $SL(2,Z)$ and Eisenstein Functions} 

In this section we describe the generalized Eisenstein series on the 
fundamental domain of $U(1)\backslash SL(2,R)/ SL(2,Z)$ and their relevant 
properties.  These functions are non-holomorphic modular forms defined by 

\bqr  
E_s^{(q,-q)}(\tau,\bar\tau) = \sum_{(m,n)\neq (0,0)} {\tau_2^s\over 
(m+n\tau)^{s+q} (m+n\bar\tau)^{s-q}}  \ ,
\label{modular}
\fqr 
where the sum over the integral pairs $(m,n)$ does not include $m=n=0$. For 
$q=0$ we recover the non-holomorphic Eisenstein series.  The series converges 
for $s>1$ on the fundamental domain ($\vert\tau_1\vert \leq 1/2$ and 
$\vert \tau\vert \geq 1$) and transforms under the fractional linear 
transformation 
\bq  
\tau \rightarrow \tau'={a\tau+b\over c\tau+d} \ ,   
\fq 
as 
\bqr  
E_s^{(q,-q)}(\tau',\bar\tau') = \Bigl[{c\tau+d\over c\bar\tau+d}\Bigr]^q ~
E_s^{(q,-q)}(\tau,\bar\tau) \ .  
\fqr 
Although the Eisenstein series are modular invariant, the generalized series 
for $q\neq 0$ transforms with a weight $(q,-q)$.  The latter may be used to find 
additional modular invariant functions by pairing them together as in $E_s^{(q,-q)} 
(\tau,\bar\tau) E_s^{(-q,q)}(\tau,\bar\tau)$ together with additional $n$-tuple 
products where the weights add up as $\sum_{i=1}^n q_i=0$.   

The generalized Eisenstein series $E_s^{(q,-q)}(\tau,\bar\tau)$ are 
related differentially to $E_s(\tau,\bar\tau)$ by 
\bqr  
E_s^{(q+1,-q-1)}(\tau,\bar\tau) = \Bigl( i\tau_2{\partial\over\partial\tau} 
+ {q\over 2} \Bigr) E_s^{(q,-q)} \ ,  
\label{nonholrel} 
\fqr 
and 
\bqr 
E_s^{(q-1,q+1)}(\tau,\bar\tau) = \Bigl( -i\tau_2{\partial\over\partial\bar\tau} 
-i{q\over 2} \Bigr) E_s^{(q,-q)} \ .  
\fqr 
The asymptotic expansion for $q\neq 0$ may be obtained from $q=0$ via 
the relation in \rf{nonholrel}. 

Further, the functions satisfy the $SL(2,Z)$ covariantized differential 
relation, 
\bq  
4 \Bigl(\tau_2 {\partial\over\partial\tau} + i {1-q\over 2} \Bigr)  
\Bigl( \tau_2 {\partial\over\partial\bar\tau} - i{q\over 2} \Bigr)  
E_s^{(q,-q)}(\tau,\bar\tau) = \lambda_s^{(q,-q)}  
 E_s^{(q,-q)}(\tau,\bar\tau) \ , 
\fq 
with an iteratively constructed eigenvalue from \rf{nonholrel}.  

The complete asymptotic form for large $\tau_2$ is found via manipulating 
the series through a Poisson resummation.  The general form is 
\bqr  
E_s^{(0,0)} (\tau,\bar\tau) = a_s \tau_2^s + b_s \tau_2^{1-s}  
\fqr 
\bqr 
+  {2\sqrt{\tau_2} \pi^s\over \Gamma(s)} \sum_{(m,n)\neq (0,0)} 
 \vert{m\over n}\vert^{s-{1/2}} K_{s-{1\over 2}} (2\pi\tau_2 \vert mn\vert) 
 e^{2\pi imn\tau_1} \ ,  
\label{modexpand}
\fqr 
where $K_s(x)$ is the standard modified Bessel function with expansion 
for large $x$, 
\bqr 
K_r(x) = ({\pi\over 2x})^{1\over 2} e^{-x} ~\Bigl[ \sum_{n=0}^\infty {1\over 
 (2x)^n} {\Gamma(r+n+{1\over 2}) \over \Gamma(r-n+{1\over 2}) \Gamma(n+1)}  
 \Bigr] 
\fqr 
and where the coefficients in \rf{modexpand} are,  
\bqr  
a_s = 2 \zeta(2s) \qquad\qquad b_s = 2\sqrt{\pi} \zeta(2s-1) 
 {\Gamma(s-{1\over 2})\over \Gamma(s)} \ .  
\fqr 
The asymptotic form for all $s>1$ contains two terms which are powers of 
$\tau_2$ together with an infinite series of exponentially suppressed 
terms.  The former will relate to the perturbative expansion of the 
superstring S-matrix and the latter to a series of conjectured D-instanton 
corrections in the uncompactified theory.  The generalized Eisenstein functions 
also have the same structure; however, the coefficients are different.  For 
example, 
\bqr  
a_s^{(1,-1)} = 2 s\zeta(2s)   \qquad\qquad  b_s^{(1,-1)} = 2\sqrt{\pi} (1-s) 
 \zeta(2s-1) {\Gamma(s-{1\over 2})\over \Gamma(s)}   \ .
\fqr 
All of these modular forms for general $q$-values give rise to two power 
suppressed terms in the large $\tau_2$ limit together with exponentially 
suppressed corrections.  

\section{U-dualities and Constraints} 

In order to examine the same construction of the S-matrix 
in dimensions other then ten and compare with maximally extended 
supergravity theories, we shall compactify IIB superstring 
theory on $M_d\times T^{10-d}$.  In this section, we examine the 
modular form construction and find the similar truncation property 
to $d=10$ in perturbation theory.  

The S-duality inherited in ten dimensions is modified to the discrete 
U-duality group acting on all of the moduli of the toroidally 
compactified theory \cite{Hull:1995ys}, and the low-energy 
theory is described by the dimensional reduction of the $d=10$ IIB 
supergravity.  The non-perturbative vacuum is parameterized by the full 
complement of scalar fields in the lower dimensional maximally extended 
supergravity.   In this section, we generalize the previous construction for 
the S-matrix of ten-dimensional string theory to the compactified 
cases with the use of additional automorphic functions.  

\begin{table} 
\caption{The U-duality structure in toroidally compactified 
IIB superstring theory.  The $Z_2$ in $d=9$ reflects the element 
that takes the IIB string into the IIA.} 
\begin{center}
\begin{tabular}{|l|c|} \hline 
{\em } & {\rm Non-perturbative U-Duality Group}  \\ \hline 
$d=10$ &  $SL(2,Z)$  \\ \hline  
$d=9$  &  $SL(2,Z) (\times Z_2) $  \\ \hline  
$d=8$  &  $SL(2,Z)\times SL(3,Z)=E_{3(3)}(Z)$  \\ \hline 
$d=7$  &  $SL(5,Z)=E_{4(4)}(Z)$  \\ \hline 
$d=6$  &  $SO(5,5,Z)=E_{5(5)}(Z) $ \\ \hline
$d=5$  &  $E_{6(6)}(Z)$  \\ \hline 
$d=4$  &  $E_{7(7)}(Z)$  \\ \hline   
\end{tabular}
\end{center}   
\end{table}

The vacuum state of compactified supergravity is parameterized by 
the values of scalars living in the symmetric space $G(R)/H(R)$ where 
$H$ is the maximal compact subgroup of $G$; we shall take the 
group $G(R)=E_{p+1(p+1)}(R)$ which contains as a subgroup $SO(p,p,R)$ (with 
the latter having maximal compact subgroup $SO(p,R)\times SO(p,R)$.  
Duality transformations 
generate an equivalence class of theories identified by an action of 
the infinite discrete group $G(Z)$.  In compactifications $M_d\times 
T^{10-d}$ there is a T-duality group $SO(d,d,Z)$ and a non-perturbative 
action of $SL(2,Z)$ (the coupling constant of the dilaton and axion scalar 
of the uncompactified IIB superstring) inherited from the uncompactified 
ten-dimensional IIB superstring.  The full symmetry of the equations of 
motion and the quantum U-duality group is known to be larger and is listed 
in table 5 \cite{Cremmer:1981zg}.  (We will not consider the $O(d,d,Z)$ 
enlargement of the T-duality group as the elements with negative determinant 
exchange the IIB string with the type IIA one.)  The larger U-duality group 
$E_{11-d(11-d)}(Z)$ ($d\leq 9$) contains the subgroup $SO(10-d,10-d,Z)\times SL(2,Z)$ 
for different $d$.  It treats all of the scalars in the supergravity on 
the same footing, although the dilaton plays a special role in the 
compactified string in measuring the loop expansion.  The general 
form of the symmetry and the constraints on the functional form of 
automorphic terms in the low-energy expansion of the S-matrix  
appear to fix the functional form of scattering in M-theory, defined in this 
case by T-dualizing the result of graviton scattering in IIB on $S_R$ and 
decompactifying the two-torus with complex structure $\tau$. 

The simplest theory from the above is uncompactified type IIB superstring 
theory which is parameterized by $U(1) \backslash SL(2,R)$.  Duality transformations 
take the theories and maps them to equivalent theories under $SL(2,Z)$; the 
fundamental domain of $U(1) \backslash SL(2,R) / SL(2,Z)$ parameterizes inequivalent 
vacua of the ten-dimensional IIB superstring.  In the compactified theories 
the inequivalent theories are described 
by vacuum expectation values of the set of moduli parameterizing the fundamental 
domain of $H(R) \backslash E_{11-d(11-d)}(R) /~ E_{11-d(11-d)}(Z)$.  In analogy with 
the ten-dimensional case we consider as building 
blocks of the S-matrix the $E_{11-d(11-d)}(Z)$ invariant functions which satisfy 
the differential equations 
\bqr  
\nabla_{H(R) \backslash E_{11-d(11-d)}(R)} f_s(\phi_j) = \lambda_{11-d,s} f_s(\phi_j) \ , 
\label{Ediff}
\fqr 
and their generalizations, in analogy with the Eisenstein series for 
the ten-dimensional uncompactified superstring theory.  This form immediately 
decompactifies to higher dimensional U-duality invariant Eisenstein series 
differentially because the metric reduces on the U-duality subgroup.  
These functions 
relevant for the lowest derivative term, i.e. $R^4$, have been discussed 
for $d\geq 7$ in \cite{Kiritsis:1997em} and for general integer dimensions 
in \cite{Obers:1999um}. 
In the latter work the unified set of perturbative and non-perturbative 
contributions contributions to the $R^4$ term was argued to be described 
by 
\bqr 
A_d = {1\over \kappa_d^2} \int d^dx \sqrt{g} ~f_d(\phi_j) R^4 \ ,
\fqr 
where,  
\bqr  
f_d(\phi_j) = E_{\rm string;s=3/2}^{E_{11-d(11-d)}(Z)} (\phi_j) \ , 
\label{ffunctddim}
\fqr 
where $\kappa_d^2$ is the gravitational coupling constant in $10-d$ dimensions 
and the Eisenstein function takes into account the summation over the 
``string multiplets''; the string multiplet is a unified representation 
of the full duality group $E_{11-d(11-d)}(Z)$ of the particle and membrane 
representations with given charges under the T-duality group and we refer 
the reader to \cite{Obers:1999um} for the construction of the general 
invariants \rf{ffunctddim}.  This 
function gives rise to perturbative contributions at tree-level 
and one-loop only which is in accord with the explicit results of 
maximal supergravity in $d$ dimensions in \cite{Bern:1998ug}.    
It also limits consistently to reproduce decompactified dimensions 
as well as giving agreement with the known tree- and one-loop 
level contributions.  The perturbative truncation property of the 
function in \rf{ffunctddim} permits contributions at only
tree and one-loop level and is consistent with the tensor properties 
of maximally extended supergravity in different integer dimensions 
\cite{Bern:1998ug}.

The general constrained Eisenstein functions for a symmetric space $G(R)/H(R)$ 
are described in detail in \cite{Obers:1999um}, and are defined 
for a given representation ${\cal R}$ of $G$ by, 
\bqr  
E_{R,s}^{G(Z)}(\phi_j) = \sum_{m\in \Lambda_R \neq 0} {\delta(m\wedge m) 
\over \left[ M^2(m) \right]^s} \ .  
\label{Emoddef} 
\fqr 
The form in \rf{Emoddef} is analogous to constructing modular invariant 
functions on the fundamental domain of $SL(2,Z)$ in the manner of 
\bqr  
g(\tau) = \sum_{\gamma\in G} f(\gamma\cdot \tau) \ , 
\fqr 
but with additional scalars involved in the mass formula in the 
denominator of \rf{Emoddef}.  

In \rf{Emoddef}, $\phi_j$ denotes elements in $G(R)/H(R)$ and $m$ is a vector 
in the integer lattice $\Lambda_R$ transforming in the representation 
$R$.  The $m\wedge m =0$ condition projects onto sets of integers $m$ 
so that the symmetric tensor product $R\times_s R$ gives its highest irreducible 
multiplet which defines $m$ to be the most symmetric piece of the 
direct product.  Physically, the set of integers in the lattice 
$\Lambda_R$ labels the set of BPS states in the representation $R$ of 
the duality group (for example, one may consider the unified 
string representation of $E_{11-d(11-d)}(Z)$ or its further decompositions.)  
The condition $m\wedge m=0$ is the half BPS condition, and 
the mass formula contributing to the sum in the Eisenstein series 
for sets of states contributing to \rf{Emoddef} is 
\bqr  
M^2_{\rm BPS}(m) = m \cdot M\cdot m \ .  
\fqr 
One could relax the condition $m\wedge m=0$ to include quarter 
and eigth BPS states by considering a more general definition of the 
the constrained Eisenstein function in \rf{Emoddef}.  The functions $E_R^{G(Z)}(\phi_j)$ 
are by construction invariant under the duality group when $G(Z)=E_{11-d(11-d)}(Z)$ and 
take values in the fundamental domain $H(R) \backslash E_{11-d(11-d)}(R) / 
E_{11-d(11-d)}(Z)$.   The various 
discrete U-duality groups are listed in table 4.   

In analogy with the generalized Eisenstein functions on  
the $U(1) \backslash SL(2,R) / SL(2,Z)$ fundamental domain, a  
generalized series $E^{(q,-q),E_{11-d(11-d)(Z)}}_{R,s} (\tau,\bar\tau)$ 
in the limit of zero moduli may be given.  The asymptotic form 
in \rf{Easymp} may be given a weight by covariantizing as in \rf{nonholrel}.  
The coupling $\tau_2 V_{10-d}^{4\over 10-d}$ is inert under $SL(2,Z)$ 
transformations.  

The simplest forms in the $d$-dimensional set are the Eisenstein functions 
relevant to uncompactified IIB superstring theory.  The construction 
given in previous sections generalizes similarly, although the precise 
functional form is more complicated, to the toroidally compactified cases; 
however, the $E_{11-d(11-d)}(Z)$ Eiseinstein functions relevant to lower target 
dimensions have different convergence properties but the same type of 
perturbative truncation to a finite order in the coupling constants as is 
found in later sections in the small volume and null moduli limit where only 
the action on the $SL(2,Z)$ transformations is taken on the string coupling.  
As a consistency in the decompactification limit where 
the radii are taken to infinity, the successively higher-dimensional 
modular ans\"atze for the S-matrix should be retrieved.   

In order to generalize the S-matrix form 
in dimensions $d\geq 7$ analogous to the one in ten dimensions, 
we need a systematic treatment of the Eisenstein series only for the 
$SL(p\leq 5)$ groups, while in lower dimensions the exceptional groups 
enter as special cases with representations that may be decomposed into 
the $SL(p)$ ones or the T-duality ones.  

The symmetry of the supergravity field equations pertaining to the 
U-duality group enforce that the different states and solitonic 
configurations from the compactified string fall into representations 
of the duality group \cite{Hull:1995ys}.  These representations 
correspond to half-BPS states and are used in the construction 
\rf{Emoddef} to find invariant functions under $E_{11-d(11-d)}(Z)$ which 
live on the fundamental domain parameterizing the moduli space of 
the compactified IIB superstring theory (listed in Table 3).  As discussed 
in \cite{Obers:1999um}, these representations are listed in tables 4-6, and 
there may be degeneracy amongst the different constrained Eisenstein 
series between the functions constructed from the different representations 
at lowest order in derivatives, $R^4$.  Consistency requires the higher-derivative 
terms in the low-energy limit to reduce under 
decompactification to the higher-dimensional forms (in integer dimensions).  

\hskip -.3in
\begin{table} 
{\small
\caption{``String'' multiplets of $E_{11-d(11-d)}$ relevant 
to compactified $M_d\times 
T^{10-d}$ IIB superstring theory and their decompositions into  
$SL(11-d,Z)$ and T-duality $SO(10-d,10-d)$ groups.}
\begin{center}
\begin{tabular}{|l|c|c|c|c|} \hline 
{\em d} & {\rm U-Duality Group} & string rep & $SL(11-d)$ rep 
 & SO(10-d,10-d) rep \\ \hline 
$10$ &  $SL(2,Z)$ & {\bf 1} & {\bf 1} & {\bf 1}  \\ \hline  
$9$  &  $SL(2,Z)$ & {\bf 2} & {\bf 2} & 
  {\bf 1} + {\bf 1} \\ \hline  
$8$  &  $SL(2,Z)\times SL(3,Z)$ & {\bf (1,3)} & {\bf 3} & 
   {\bf 1} + {\bf 2}  \\ \hline 
$7$  &  $SL(5,Z)$ & {\bf 5} & {\bf 4}+{\bf 1} & 
 {\bf 4} + {\bf 1} \\ \hline 
$6$  &  $SO(5,5,Z)$ & {\bf 10} & {\bf 5}+{\bf $\bar 5$} & 
 {\bf 1} + {\bf $8_S$} + {\bf 1} \\ \hline 
$5$  &  $E_{6(6)}(Z)$ & {\bf ${\bar 27}$} & {\bf 6} + {\bf 15} + 
 {\bf $\bar 6$} & {\bf 1} + {\bf 10} + {\bf 16}  \\ \hline 
$4$  &  $E_{7(7)}(Z)$ & {\bf 133} & {\bf 7} + {\bf 28} + {\bf 35} + $\ldots$ & 
 {\bf 1} + ({\bf 1}+ {\bf 66}) + {\bf 32} + $\ldots$ \\ \hline  
\end{tabular} 
\end{center} }
\end{table}

In the remainder of the section we confirm the $\tau_2$ 
dependence in Einstein frame of the perturbative series in different 
integral dimensions through comparison with the expansions of 
the appropriate modular forms at small coupling.  The agreement of 
the expansions with the powers of $\tau_2$ is a check on the modular 
properties of the scattering in different dimensions.  

The transformation of string frame to Einstein frame is given by 
\bqr  
g_{\mu\nu}^{(s)} = e^{\alpha \phi} g_{\mu\nu}^{(E)}  \ , 
\fqr 
together with 
\bqr  
\sqrt{g_{(s)}} = e^{\alpha {d\over 2}\phi} \sqrt{g_{(E)}} \qquad \qquad 
 R_{(s)} = e^{-\alpha\phi} R_{(E)} \ .  
\fqr 
The Einstein frame of the Einstein-Hilbert action derived from the 
string frame is found from, 
\bqr  
\alpha = {4\over d-2} \ , 
\fqr 
or in terms of the ten-dimensional string coupling $g_s^{(10)}$ 
\bqr  
g^{(s)}_{\mu\nu} = \tau_2^{-{4\over d-2}} g^{(E)}_{\mu\nu} \qquad 
s_{ij}^{(s)} = \tau_2^{4\over d-2} s^{(E)}_{ij} \ , 
\label{transf} 
\fqr  
where we have illustrated how invariants scale also in \rf{transf}.  
 
\begin{table} 
\caption{``Particle'' multiplets of $E_{11-d(11-d)}$ relevant to 
compactified $M_d\times T^{10-d}$ IIB superstring theory and 
their decompositions into the fundamental $SL$ and T-duality 
$SO(10-d,10-d)$ groups.}    
\begin{center} 
{\small
\begin{tabular}{|l|c|c|c|c|} \hline 
{\em d} & {\rm U-Duality Group} & particle rep & SL rep 
 & SO rep \\ \hline 
$10$ &  $SL(2,Z)$ & {\bf 1} & {\bf 1} & {\bf 1}  \\ \hline  
$9$  &  $SL(2,Z) (\times Z_2) $ & {\bf 3} & {\bf 1}+ {\bf 2} & 
  {\bf 1} + {\bf 2} \\ \hline  
$8$  &  $SL(2,Z)\times SL(3,Z)$ & {\bf (2,3)} & {\bf 3}+{\bf $\bar 3$} & 
   {\bf 2} + {\bf 4}  \\ \hline 
$7$  &  $SL(5,Z)$ & {\bf 10} & {\bf $\bar 4$}+{\bf 6} & 
 {\bf 4} + {\bf 6} \\ \hline 
$6$  &  $SO(5,5,Z)$ & {\bf 16} & {\bf $\bar 5$}+{\bf 10}+ {\bf 1} & 
 {\bf $8_v$} + {\bf $8_c$} \\ \hline 
$5$  &  $E_{6(6)}(Z)$ & {\bf 27} & {\bf 6} + {\bf 15} + 
 {\bf $\bar 6$} & {\bf 1} + {\bf 10} + {\bf 16}  \\ \hline 
$4$  &  $E_{7(7)}(Z)$ & {\bf 56} & {\bf 7} + {\bf 21}+ {\bf $\bar 7$} + 
 {\bf ${\bar 21}$} & {\bf 12} + {\bf 32} + {\bf 12} \\ \hline  
\end{tabular} } 
\end{center} 
\end{table}

In \rf{transf} we see that dimensions $d< 6$ is special: The Einstein frame 
coupling dependence increases by $\tau_2^\beta$ with $\beta>1$ 
for every pair of derivatives in the low-energy expansion and the 
transition dimension is $d=6$.  Six dimensions is also singled out due 
to the presence of the four-form self-dual moduli.  At weak IIB coupling, 
$\tau_2\rightarrow 
\infty$, the dimensions $d\leq 6$ have the feature that the tree-level 
terms are larger in string coupling at higher derivatives (in $d>6$ the 
terms are of smaller value), although the appropriate 
expansion parameter in this comparison at low-energy is 
\bqr  
\alpha = p^2_{(E)} \tau_2^{4\over d-2} \ ,  
\fqr 
which may give a decreasing effect if the momentum scale is such 
that $\alpha <1$.  (The volume modulus is given in \rf{treekform}).  
The coupling constant dependence in the low-energy S-matrix expansion 
has the distinguishing feature in $d\leq 6$, the same dimensions in 
which supergravity has extra finiteness properties in the Regge limit 
and also to higher loop orders implied by the automorphic IIB graviton 
scattering.  

\hskip -.3in
\begin{table} 
{\small 
\caption{
``Membrane'' multiplets of $E_{11-d(11-d)}$ relevant to 
compactified $M_d\times 
T^{10-d}$ IIB superstring theory and their decompositions into the mapping 
class $SL(10-d)$ and T-duality $SO(10-d,10-d)$ groups.} 
\begin{center}
\begin{tabular}{|l|c|c|c|c|} \hline 
{\em d} & {\rm U-Duality Group} & memb. rep & SL rep 
 & SO rep \\ \hline 
$10$ &  $SL(2,Z)$ & {\bf 1} & {\bf 1} & {\bf 1}  \\ \hline  
$9$  &  $SL(2,Z) $ & {\bf 1} & {\bf 1}  & 
  {\bf 1} \\ \hline  
$8$  &  $SL(2,Z)\times SL(3,Z)$ & {\bf (2,1)} & {\bf 1}+{\bf 1} & 
   {\bf 2}  \\ \hline 
$7$  &  $SL(5,Z)$ & {\bf $\bar 5$} & {\bf 4}+{\bf 1} & 
 {\bf 4} + {\bf 1} \\ \hline 
$6$  &  $SO(5,5,Z)$ & {\bf ${\bar 16}$} & {\bf 5}+{\bf 10}+ {\bf 1} & 
 {\bf $8_v$} + {\bf $8_c$} \\ \hline 
$5$  &  $E_{6(6)}(Z)$ & {\bf 78} & {\bf 1} + {\bf 20} + 
 {\bf 36} + $\ldots$ & {\bf 16} + {\bf $\bar 16$} + ({\bf 1}+{\bf 45}) \\ \hline 
$4$  &  $E_{7(7)}(Z)$ & {\bf 912} & {\bf 1} + {\bf 35}+$\ldots$ & 
  {\bf 32} + $\ldots$ \\ \hline  
\end{tabular} 
\end{center} } 
\end{table} 

Alternatively, we may transform to Einstein frame in ten dimensions 
and then compactify on an $T^{10-d}$ torus, taking into account the 
volume dependence of the torii.  The frame dependence in \rf{transf}  
forces the perturbative series in Einstein frame of the compactified 
IIB superstring to have tree-level contributions to $\Box^k R^4$ with 
the $\tau_2$ dependence,   
\bqr  
A_4^{\rm tree,k} = N_k \Bigl[ \tau_2 V_{10-d}^{4\over 10-d} \Bigr]^{ 
 ({3+k\over d-2})({10-d\over 2})} \tau_2^{{3\over 2} + {k\over 2}} 
 \int d^dx \sqrt{g_{E}} ~ \Box^k_{(E)} R^4_{(E)}  
\fqr 
\bqr 
= N_k V_{10-d}^{2(3+k)\over (d-2)} 
   \tau_2^{12+4k\over d-2} \int d^dx \sqrt{g_{E}} ~ \Box^k_{(E)} 
 R^4_{(E)} \ ,  
\label{treekform} 
\fqr 
where the volume factor $\tau_2 V_{10-d}^{4\over 10-d}$ is inert under 
the subgroup $SL(2,Z)$ transformations of the U-duality group in Einstein frame.  
In $d=10$ the perturbative series reproduces the series in \rf{kform} and sub-leading 
string genus corrections to \rf{treekform} generate powers $\tau_2^{-2g}$ 
relative to the tree-level; the coefficient is modified in the toroidal 
compactifications due to the volume dependence and scaling in $d$-dimensional 
compactified Einstein frame.  We collect some explicit powers of $\tau_2$ 
dependence of the $\Box^k R^4$ terms in the low-energy expansion of the 
IIB superstring S-matrix on $M_d\times T^{10-d}$ dimensions below: 

\begin{table} 
\caption{Example couplings in $\tau_2^{\beta}$ at genus zero in 
Einstein frame.} 
\begin{center}
\begin{tabular}{|l|c|c|c|c|} \hline 
{\em d} & $R^4$ & $\Box R^4$  & $\Box^2 R^4$  & $\ldots$ \\ \hline 
$10$ & ${3\over 2}$ & $2$  & ${5\over 2}$  &  \\ \hline  
$9$  & ${12\over 7}$  & ${16\over 7}$ & ${20\over 7}$ &  \\ \hline  
$8$  & $2$  & ${10\over 3}$ & ${14\over 3}$ &  \\ \hline 
$7$  & ${12\over 5}$ & ${16\over 5}$ & $4$ &  \\ \hline 
$6$  & $3$  & $4$ & $5$ &  \\ \hline 
$5$  & $4$  & ${16\over 3}$ & ${20\over 3}$  &  \\ \hline 
$4$  & $6$  & $8$  & $10$ &  \\ \hline  
\end{tabular} 
\end{center}  
\end{table}

The $\tau_2$ dependence in \rf{treekform} at tree-level in Einstein 
frame may be matched with the modular form expansions in $d\leq 10$ 
in accord with criteria (1) in section 2.  The Eisenstein function 
in lower dimensions on the appropriate fundamental domain with 
value $s=3/2$ and $s=5/2$ gives rise to the dependence in Table 7.  

\subsection{Asymptotic Limits} 

We give the large coupling expansions in this subsection and first examine 
the simplest case, namely the $SL(n,Z)$ Eisenstein functions to the 
fundamental (and anti-fundamental) representations.  The $SL(2,Z)$ 
Eisenstein function to the fundamental representation is given by 
\bqr  
E_s(\tau,\bar\tau) = \sum_{(p,q)\neq (0,0)} {\tau_2^s \over \vert 
p+q\tau\vert^s} \ , 
\fqr 
and for $SL(n,Z)$ in the fundamental representation we have the 
explicit analogous function, 
\bqr  
E^{\rm SL(n,Z)}_{R=d,s}(\phi_j) = \sum_{(m^1,\ldots,m^n)\neq 0} 
 {1\over \left[ m^i g_{ij}(\phi_k) m^j \right]^s }  \ . 
\label{fundEgen}  
\fqr 
The anti-fundamental representation is given by \rf{fundEgen} but with 
indices lowered/raised.  The series in \rf{fundEgen} is absolutely 
convergent only for $s>n/2$.  

Upon taking a $d$-dimensional torus into a direct product 
of one circle of radius $R$ times a $T^{d-1}$ generically non-orthogonal 
torus together with large $R$, the limit of \rf{fundEgen} is 
\bqr  
E^{\rm SL(d,Z)}_{R=d,s}(\phi_j) = E^{\rm SL(d-1,Z)}_{R=d-1,s}(\tilde\phi_j)
+ {2\pi^s\Gamma(s-{d-1\over 2}) \zeta(2s-d+1) \over \pi^{s-{d-1\over 2}} 
\Gamma(s) R^{2s-d+1} V_{d-1} }  
\fqr 
\bqr  
+ {2\pi^s\over \Gamma(s) R^{2s-d-1}} \sum_{m^a,n^b} \vert { n^a g_{ab} n^b \over 
 m^2} \vert K_{s-{d-1\over 2}}\Bigl(  2\pi \vert m\vert R \sqrt{n^a g_{ab} n^b} 
 \Bigr) \ .  
\label{asymptSLd}
\fqr 
The asymptotic limit in \rf{asymptSLd} of $E^{\rm SL(d,Z)}_{R=d,s}(\phi_j)$ 
gives rise to the Eisenstein series of $SL(d-1)$ to the fundamental 
representation together with an infinite number of terms depending on 
the large radius (one power-behaved and the rest exponentially suppressed) 
which vanish as $R\rightarrow\infty$.  Although only the fundamental representation 
is analyzed in \rf{asymptSLd} a similar structure is expected for other representations 
of the duality group $U(Z)$; the zero moduli approximation also generates this 
limit.  This structure is consistent with 
the expected form of the modular properties of the S-matrix in the 
decompactification limit and the exponential terms in the expansion of the 
forms represent the various 
wrapping of string solitonic states in the compact directions of the 
compactified torii dimensions.  In subsequent work we shall analyze the 
$d$-dimensional form in view of the finite radii dependence of the toroidal 
compactifications.  

The weak coupling expansion is defined by taking the  
dilaton IIB scalar to large vacuum expectation value (small string $g_s$ coupling 
constant).  Turning off all of the vacuum values 
of the scalar fields except for $\tau$ which is acted on by the S-duality $SL(2,Z)$ 
subgroup of $U(Z)$, these (regulated) functions reduce to the Eisenstein series for 
$E^{\rm SL(2,Z)}_{s}(\tau,\bar\tau)$ together with a perturbative truncation 
property described in previous sections that gives rise to the same power counting 
in the low-energy limit of the superstring scattering elements.

We can consider the special point of the moduli of the toroidally compactified 
theory when all moduli except for the ten-dimensional dilaton have vanishing 
values.  In this case, the large $\tau_2$ behavior of the Eisenstein 
series in \rf{Ediff} may be constructed by solving the differential equation 
for the $SL(p,Z)$ representation decompositions into the $SL(2,Z)$ 
subgroup of the various string multiplet contributions.  The functional 
form of the small moduli contribution is invariant under $SL(2,Z)$ 
transformations and is,   
\bqr  
E_s^{E_{11-d(11-d)}} (\tau,\bar\tau) = \Bigl[ \tau_2 V_{10-d}^{4\over 10-d} 
\Bigr]^{ 
 ({3+k\over d-2})({10-d\over 2})}~ \sum_{(p,q)\neq (0,0)} 
 {\tau_2^s\over \vert p+q\tau\vert^{2s}} + \ldots \ ,
\label{Easymp}
\fqr 
for $k=2s-3$, which truncates in every dimension, as it does in $d=10$ 
(for the subtracted or convergent series $E_s$).  The form in \rf{Easymp} 
is due to the $SL(2,Z)$ subgroup of the larger U-duality group.  Constructions 
using the generalized 
Eisenstein series follows similarly in the toroidal compactifications, 
with the difference involving the volume factor of the $10-d$ dimensional 
tori.  The unitarity construction described in previous sections gaurantees 
that the imaginary parts will also be invariant under the U-duality group, 
after construction of the supersymmetric extension of the $\Box^k R^4$ 
terms and integrating to find the imaginary parts.  

\subsection{D=8 Example}   

In the remainder of this section we give an example for the polynomial 
terms in the low-energy expansion of the S-matrix in the compactified 
theories analogous to the one presented for the uncompactified IIB 
superstring but invariant under the larger U-duality group.  We give the 
example of the $D=8$ theory which has a U-duality group in the toroidally 
compactified theory of $SL(2,Z)\times SL(3,Z)$ and a set of scalars 
parameterizing the fundamental domain of $SO(2,R)\times SO(3,R) \backslash 
SL(2,R)\times SL(3,R) / SL(2,Z)\times SL(3,Z)$.  An explicit parameterization 
of the moduli space consists of the complex structure of the two-torus 
with metric (and volume $V=V_2$), 
\bqr  
g_{ab}= {V\over U_2} 
\left( \begin{array}{cc} 
  1 & U_1 \\ 
  U_1 & \vert U\vert^2 \end{array} \right)
\label{2torusmetric}
\fqr 
where $U$ spans the region $U(1)\backslash SL(2,R)$ and with the enhanced 
moduli of $SL(3,R)/SO(3)$.  The scalar $\tau_2 V$ is invariant under 
the $SL(2,Z)$ sub-group of $SL(3,Z)$ acting on the IIB scalar $\tau$.  
The moduli of $SL(3,R)/SO(3)$ consist of the dilatonic 
and axionic couplings $\tau$ together with those of the reduction of the 
anti-symmetric complex tensors $B_{NS,R}=\epsilon_{ab} B_{NS,R}^{ab}$ grouped 
as $B=B_R+\tau B_{NS}$ into the symmetric coset matrix form with the 
matrix satisfying the determinant ${\rm det}M=1$ condition, 
\bqr  
M= (\tau_2 V)^{-1/3} \left( \begin{array}{ccc}  
 {1\over \tau_2} & {\tau_1\over\tau_2} & {{\cal R}(B)\over \tau_2} \\ 
 {\tau_1\over\tau_2} & {\vert \tau\vert^2\over \tau_2} & {{\cal R}({\bar \tau} B) 
 \over \tau_2} \\ 
 {{\cal R}(B)\over \tau_2} & {{\cal R}({\bar \tau} B)  \over \tau_2}
 & \tau_2 V^2 + {\vert B\vert^2 \over \tau_2} \end{array} \right) \ .
\label{d8param}
\fqr 
The matrix $M$ in \rf{d8param} together with $U$ parameterizes the 
scalar manifold $U(1)\backslash SL(2,R) \times SL(3,Z) / SO(3,R)$ of the 
classical IIB superstring theory compactified on a two-torus with metric 
in \rf{2torusmetric} and the complex structure parameterized by $U$. 
The action of the scalars and Einstein-Hilbert low-energy theory is 
\bqr  
S = {1\over \kappa_8^2} \int d^8x ~\sqrt{g} \Bigl[ R 
 - {\partial U \partial {\bar U} \over U_2^2} 
 +{1\over 4} {\rm Tr}\bigl( \partial M \partial M^{-1}\bigr) \Bigr] \ ,
\fqr 
and we are primarily interested in this work in the string 
corrections to gravitational action.  The Eisenstein series relevant 
to the eight-derivative term has been elucidated for the $R^4$ term in 
\cite{Kiritsis:1997em,Obers:1999um}.  

The $R^4$ term in Einstein frame arising in the low-energy expansion 
is described in \rf{ffunctddim}.  In accord with the functional form 
in \rf{twotermspoly} and \rf{tendimset} we take the twelve derivative 
term to be 
\bqr  
S^{12}_{d=8} = \int d^8x \sqrt{g} 
 ~ E_{\rm string;s=5/2}^{SL(2,Z)\times SL(3,Z)} (\phi_j) ~ \Box^2 R^4 \ , 
\label{d8twelveder}
\fqr  
which generalizes to arbitrary dimensions to 
\bqr  
S^{12}_{d}= \int d^dx \sqrt{g}~ E_{\rm string;s=5/2}^{E_{11-d(11-d)}(Z)} (\phi_j) 
 ~ \Box^2 R^4 \ . 
\fqr 
The string multiplet in $d=8$ decomposes under $SL(2,R)\times SL(3,Z)$ 
into the representation ${\bf 1}\times {\bf 3}$, and accordingly, the 
automorphic contribution in \rf{d8twelveder} breaks into, 
\bqr  
S^{12}_{d=8} = \int d^8x \sqrt{g} ~ 
 E_{{\bf 3};5/2}^{SL(3,Z)} (\phi^{(2)}_j) \Bigr] ~ \Box^2 R^4 \ , 
\label{d8sltwelveform}
\fqr 
or alternatively, 
\bqr  
S^{12}_{d=8} = \int d^8x \sqrt{g} ~ \Bigl[ E_{{\bf 1};{\tilde s}_1}^{SO(2,2,Z)} 
(\phi^{(1)}_j) + E_{{\bf 2};{\tilde s}_2}^{SO(2,2,Z)} (\phi^{(2)}_j) 
 \Bigr] ~ \Box^2 R^4 \ .  
\fqr 
The modular weights $s_1=3/2$ and $s_2=5/2$ are determined by decomposing 
the form of the modular summand into the lower representations.  The 
complex structure of the torus $U$ parameterizes the $SL(2,R)/U(1)$ 
portion of the T-duality group ($SO(2,2,R)=SL(2,R)\times SL(2,R)$) and 
in the decompactification limit.

The $SL(3,Z)$ Eisenstein functions entering into \rf{d8sltwelveform} 
are explicitly in the fundamental representation constructed from 
the matrix in \rf{d8param}, 
\bqr  
E_{\bf 3;s}^{SL(3,Z)}(\phi_j^{(2)}) &=& \sum_{a,b=1,2,3} \sum_{n^a,n^b} {1\over 
 \left[ n^a M_{ab} n^b\right]^s}  
\cr && 
 = \sum_{(n^1,n^2,n^3)\neq (0,0,0)} \tau_2^s {\left(\tau_2 V\right)^{s/3} 
\over \left[ \vert n^1 + n^2\tau + n^3 B\vert^2  
 + (n^3 \tau_2 V)^2 \right]^s  } \ , 
\label{fundsl3eis}  
\fqr 
togther with the (subtracted) form of the order one Eisenstein 
funtion of the complex structure $U$, 
\bqr  
{\hat E}_{{\bf 1};1}^{SL(2,Z)} (\phi^{(1)}_j) = -\pi 
\ln\left( U\vert\eta(U)\vert^4\right) \ ,   
\fqr 
that we must also add in.  
The series in \rf{fundsl3eis} is explicitly invariant under the 
$SL(3,Z)$ transformations and to leading order in $\tau_2$ has 
the expected power dependence to agree with the Einstein frame 
string coupling at tree-level (as it must if the compactified 
string theory possesses a U-duality structure).  At small volume 
and at zero $B$ moduli we have, 
\bqr  
E_{\bf 3;{5\over 2}}^{SL(3,Z)}(\phi_j^{(2)}) &=& \tau_2^{10\over 3} V^{5\over 3} 
\sum_{(m,n)\neq (0,0)} {1\over \vert m+n\tau\vert^5} 
\cr && 
= \tau_2^{5\over 6} V^{5\over 3} \Bigl( \tau_2^{5\over 2} \zeta(5) 
+ a \tau_2^{-{3\over 2}} + {\cal O}\left({e^{2\pi i\tau}}\right) \Bigr) \ . 
\label{d8box2exp}
\fqr     
The perturbative series in \rf{d8box2exp} only receives contributions 
at genus zero and two in this limit.

In the decompactification limit $R_1,R_2\rightarrow \infty$, the 
functional form in \rf{d8twelveder} must reconcile with the 
$d=9$ and $d=10$ results for the twelve-derivative term and be 
invariant under the U-duality group $SL(2,Z)$.  Using \rf{asymptSLd}, 
we see that the only term surviving the large R limit is the 
$SL(3,Z)$ Eisenstein function in \rf{d8sltwelveform}, which correctly 
limits to an order $s=5/2$ $SL(2,Z)$ Eisenstein function.  It would 
be interesting to check the finite radius dependence further.  

\section{Perturbative truncations in integral dimensions}

In this section we examine the implications of the perturbative 
truncation, that of a maximum genus for individual terms in the 
derivative expansion, 
with maximally extended supergravity theories.  The modular structure 
of the graviton scattering amplitude indicates the perturbative truncation 
property in different dimensions and implies a similar behavior 
in supergravity.  This property is present already at order 
$\Box^2 R^4$ \cite{Green:1999pu, Green:1999pv}.  Two options to explain 
the perturbative truncation 
implied by the modular structure are either an infinite 
number of independent cancellations at each genus between the massive 
and massless modes are nullifying the contributions to $\Box^k R^4$ for 
$k\geq 2$, or that the massive modes give rise to order $\alpha'$ higher 
corrections compared with the supergravity modes in the low-energy limit 
(as at $g=0$ and $g=1$) and that IIB supergravity possesses this 
truncation.  

In order to compare the string amplitudes with 
the supergravity ones: the massive modes need to be distinguished 
from the massless ones in the $\alpha'\rightarrow 0$ limit at $g\geq 3$, 
the ghost dependence in the string amplitudes needs to be disentangled 
in the field theory limit, and regulator dependence in the supergravity 
theory implied by the integration region coming from the moduli 
space (at one-loop in \rf{g1stringinspired}) has to be examined with 
regards to possible cancellations due to the region of integration.  
Ghost dependence in the genus two four-graviton scattering amplitude, 
for example, naively indicates a contribution to the $R^4$ term in 
the derivative expansion before moduli integration that is in disagreement 
with explicit two-loop supergravity graviton scattering \cite{Iengo:1999qh}; 
consistency with S-duality and perturbative supergravity requires these 
dependencies to integrate to zero.  

In the IIB supergravity limit of the superstring scattering, primitive 
divergences in $d=10$ of the four-point amplitude are of the form: 
\bqr 
A_4^{L=1,m=0} \sim \Lambda^2 R^4 + \Box R^4 + \ldots
\fqr 
\bqr 
A_4^{L=1,m\neq 0} \sim  \Box R^4 + \Box^2 R^4 + \ldots 
\fqr 
The $\alpha'$ indicates an overall factor of $\Box$ or $s_{ij}$ in 
comparison to the massless modes after they are normalized correctly 
at the given order.

At two-loops in field theory, an explicit twelve derivatives may be 
extracted from the loop integration and the amplitude has the 
generic tensor structure, 
\bqr 
A_4^{L=2,m=0} \sim \Box^2 \left(\Lambda^6 + \Lambda^4 \Box + \Lambda^2 \Box^2 
 + \Box^3 \right) R^4 + \ldots \ .
\fqr 
Explicit string theory calculations at genus two may be 
computed in the $\alpha'\rightarrow 0$ limit.  If the massive modes 
of the string contribute at an order $\alpha'$ higher in the low-energy 
limit (as at order $g=0$ and $g=1$), then an additional pair of derivatives 
must also be extracted, 
\bqr 
A_4^{L=2,m\neq 0} \sim \Box^3 R^4 + \ldots  \ . 
\fqr 
At tree and one-loop level the massive modes explicitly contribute an 
order higher in $\alpha'$, and it is reasonable to suspect that such 
a property persists at higher order.   However, an explicit integration 
of the genus two four-point function in the low-energy limit is required; 
picture dependence in this order complicates the analysis although unitarity 
might be faulted if this were not the case.    

\begin{figure} 
\begin{center}
\epsfig{file=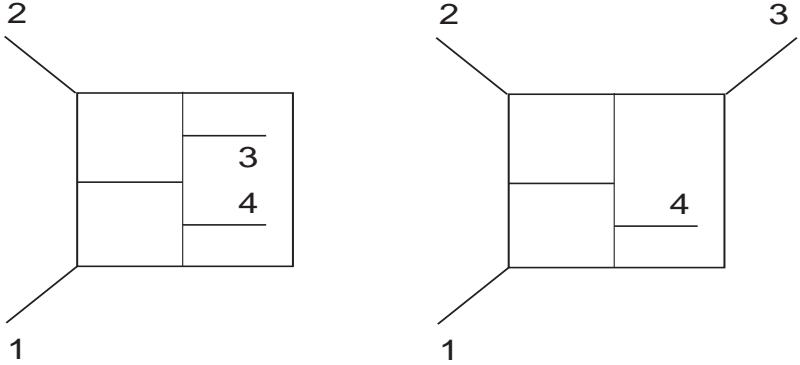,height=5cm,width=11cm} 
\end{center} 
\caption{Example non-planar Feynman diagrams in supergravity 
contributing to the four-point function computed in \cite{Bern:1998ug} illustrated 
without the additional loop tensor within the integrand.} 
\end{figure}  

At three-loops the conjectured form \cite{Bern:1998ug} of the four-point 
supergravity amplitude in ten dimensions leads to 
\bqr 
A_4^{L=3,m=0} \sim \Box^2 \left(\Lambda^{14} + \Lambda^{12} \Box + ... \right) R^4
\label{afourthree}
\fqr 
and if the massive modes contribute to a higher-order in $\alpha'$ then 
again their contribution must be of an order higher in derivatives by 
dimensional analysis, 
\bqr 
A_4^{L=3,m\neq 0} \sim \Box^3 R^4 + \ldots    \ .  
\fqr 
The contribution to $\Box^2 R^4$ in the low-energy limit of the IIB superstring 
amplitude gets contributions at most from genus two from the modular form 
$E_{5/2}(\tau,\bar\tau)$ Eisenstein function; this requires the leading 
divergence in \rf{afourthree} to have a zero coefficient. 

The agreement of the low-energy field theory modelling of the superstring 
requires an infinite number of cancellations or nullifications at 
higher genus for every $\Box^k$ term in the low-energy expansion.  The 
genus zero and two contributions to the $\Box^2 R^4$ term in the ten-dimensional 
S-matrix, for example, indicates that separately 
every contribution (primitive ones above for example) proportional 
to $\Box^2$ at $g>2$ has to be zero (every perturbative term at different 
genus differs by powers of $\tau_2^2$).  The same structure occurs because 
of the truncation $g_{\rm max}^k={1\over 2}(k+2)$ and $g_{\rm min}^k={1\over 2} 
(k+1)$, for even and odd $k$, for the higher derivatives.

The perturbative counting arising from the Eisenstein series, i.e. 
$g_{\rm min}=0$ to $g_{\rm max}^k$, is most naturally explained in the 
field theory by an explicit extraction of an additional four derivatives 
at every order (as found for example at loop order one and two).  In 
the Regge limit, where only the ladder diagrams contribute in the 
construction of \cite{Bern:1998ug}, this property is seen to infinite 
loop order for the massless modes, and pure $N=8$ supergravity at 
vanishing moduli is finite in this kinematical regime to all orders 
in four through six dimensions.  

However, at three-loops it is not clear if 
an additional set of momenta may be extracted from the complete set 
of integral functions contributing to the amplitude due to possible 
non-planar contributions possessing only three- and higher particle 
cuts; the complete form of the non-planar contributions to 
the four-point supergravity amplitude are not explicitly known to this 
order and the tensor property of three and higher loops within the 
amplitude for the supergravity modes is an open question.  Sample 
non-planar contributions are illustrated in Figure 5, where the 
internal powers of $l^4$ are not inserted \cite{Bern:1998ug}.

The tensor property, i.e. an extraction of internal loop 
momenta associated with the gravitational couplings in the pattern 
$4(1+L)$, implies that $N=8$ supergravity with all moduli 
tuned to zero except for the dilaton coupling is finite in four, 
five and six dimensions.   

The fact that the modular ansatz indicates such a cancellation in 
the different integer dimensions is suggestive of the tensor 
property, perhaps in a string-inspired regulator at higher loop 
order that preserves the S-duality structure in the supergravity 
quantum field theory.   Such cancellations after integration could 
be associated with 
boundaries of the integration region and divergence contributions 
vanishing on them in different dimensions (for example, $\sum_{(p,q)\neq 
(0,0)} 1/\vert p+q\tau\vert^4$ on the fundamental domain at genus 
one).  A complete three-loop 
supergravity divergence calculation is necessary to find out if 
there is a cancellation in the field theory (before or after 
integration), or an explicit two-loop string theory calculation to 
disentangle picture dependence in the field theory limit and 
verify the $q$ expansion of the massive modes.   

\section{Conclusions} 

We have explored a manifestly S-duality (and U-duality) invariant 
perturbative series in the derivative expansion of four-graviton 
scattering in IIB superstring theory compatible with the perturbative 
structure.   Generalizations to higher-point 
gravitational amplitudes are straightforward.  We have written the  
derivative expansion in terms of combinations of generalized Eisenstein 
series with coefficients not generally fixed by tree-level IIB graviton 
scattering.  In ten dimensions this form predicts the absence 
of singularities in the moduli space of vacua which is consistent 
with taking into account all non-perturbative corrections.  

The duality structure imposes 
strong constraints on the perturbative series from the perturbative 
superstring amplitude calculations.  Imposing the duality structure 
is an additional postulate on the S-matrix akin to unitarity in 
the approach of analytic S-matrix theory, and allows the exciting 
possibility of determining the functional form of graviton 
scattering in IIB superstring theory.   The form of the scattering 
amplitude is also useful for computing finite $\lambda= g^2N$ effects 
of $N=4$ super Yang-Mills correlation functions through the AdS/CFT 
correspondence \cite{Maldacena:1998re,Gubser:1998bc,Witten:1998qj}.  

Unitarity of the massless modes is $SL(2,Z)$ invariant by construction 
in this formulation.  The thresholds associated with the massive modes 
of the string must be found by summing the derivative expansion.  
Furthermore, these thresholds at genus one and higher in IIB string 
perturbation theory are additional constraints not yet imposed on 
the full functional form of the derivative expansion in this work, and 
could be used to fix the coefficients of the linear combinations 
of the Eisenstein series.  Factorization conditions and consistency 
of the D-instanton corrections with the coupling constant dependence 
of the instantons through a holographic AdS relation to $N=4$ super Yang-Mills 
also provide additional constraints.   

This approach, via T-duality from IIB theory on a circle, provides an 
avenue for pinning down in the derivative expansion the covariantized 
form of graviton scattering in the eleven-dimensional corner of M-theory.  
Furthermore, the same approach may be used to find the constraints of S-duality on 
the low-energy scattering obtained purely in $N=4$ super Yang-Mills 
theory, the planar form of which in the unbroken theory has been conjectured 
in perturbation theory in terms of integral functions for the four-gluon 
scattering process \cite{Bern:1997nh}.

Additional analysis through instanton calculus or constraints such 
as those listed above are necessary to determine uniqueness or possible 
contributions of cusp forms, for example, at higher derivatives which 
generate non-perturbative corrections.  Further checks of S-duality at 
the amplitude level involve low-energy expansions of $g\geq 2$ genus 
four-graviton contributions.  

The modular property in this work predicts a perturbative truncation, 
that of a maximum genus contribution for a given derivative 
order, when expanded in the string coupling 
constant beyond the known supergravity structure at two-loops.  Although 
the direct 
field theory limit of the superstring is difficult to obtain explicitly 
for genus greater than two, the most straightforward interpretation 
given the $\alpha'$ expansion in the field theory limit is 
that the perturbative series of the massless modes also truncates, 
perhaps in one regulator that preserves the remnant of the 
S- and U-duality in the supergravity approximation or field equations.  
This feature in maximally extended supergravity theory indicates 
a much higher degree of finiteness in four through six dimensions 
than expected.  

\vskip .2in
{\bf Acknowledgements}

\noi
This work is supported in part by the U.S. Department of Energy, Division 
of High Energy Physics, Contract W-31-109-ENG-38.  G.C. thanks Zvi Bern 
for numerous discussions on the finiteness properties of multi-loop 
maximally extended supergravity, and Jeffrey Harvey, Finn Larsen, Emil 
Martinec and Alan White for discussions.


\begin{thebibliography}{99}

\bibitem{Green:1997tv}
M.~B.~Green and M.~Gutperle,
Nucl.\ Phys.\  {\bf B498}, 195 (1997)
[hep-th/9701093].

\bibitem{Green:1997di}
M.~B.~Green and P.~Vanhove,
Phys.\ Lett.\  {\bf B408}, 122 (1997)
[hep-th/9704145].

\bibitem{Russo:1997mk}
J.~G.~Russo and A.~A.~Tseytlin,
Nucl.\ Phys.\  {\bf B508}, 245 (1997)
[hep-th/9707134].

\bibitem{Green:1997as}
M.~B.~Green, M.~Gutperle and P.~Vanhove,
Phys.\ Lett.\  {\bf B409}, 177 (1997)
[hep-th/9706175].

\bibitem{Green:1999pu}
M.~B.~Green, H.~Kwon and P.~Vanhove,
hep-th/9910055.

\bibitem{Berkovits:1998ex}
N.~Berkovits and C.~Vafa,
Nucl.\ Phys.\  {\bf B533}, 181 (1998)
[hep-th/9803145].

\bibitem{Townsend:1997kr}
P.~K.~Townsend,
Phys.\ Lett.\  {\bf B409}, 131 (1997)
[hep-th/9705160].

\bibitem{Cederwall:1997ts}
M.~Cederwall and P.~K.~Townsend,
JHEP {\bf 9709}, 003 (1997)
[hep-th/9709002].

\bibitem{Chalmers:1999ap}
G.~Chalmers and K.~Schalm,
JHEP {\bf 9910}, 016 (1999)
[hep-th/9909087].

\bibitem{deWit:1977fk}
B.~de Wit and D.~Z.~Freedman,
Nucl.\ Phys.\  {\bf B130}, 105 (1977).

\bibitem{Cremmer:1978km}
E.~Cremmer, B.~Julia and J.~Scherk,
Phys.\ Lett.\  {\bf B76}, 409 (1978).

\bibitem{Cremmer:1978ds}
E.~Cremmer and B.~Julia,
Phys.\ Lett.\  {\bf B80}, 48 (1978).

\bibitem{deWit:1982ig}
B.~de Wit and H.~Nicolai,
Nucl.\ Phys.\  {\bf B208}, 323 (1982).

\bibitem{Hull:1995ys}
C.~M.~Hull and P.~K.~Townsend,
Nucl.\ Phys.\  {\bf B438}, 109 (1995)
[hep-th/9410167].

\bibitem{Bern:1998ug}
Z.~Bern, L.~Dixon, D.~C.~Dunbar, M.~Perelstein and J.~S.~Rozowsky,
Nucl.\ Phys.\  {\bf B530}, 401 (1998)
[hep-th/9802162].

\bibitem{Green:1999pv}
M.~B.~Green and P.~Vanhove,
hep-th/9910056.

\bibitem{Russo:1998fi}
J.~G.~Russo,
Phys.\ Lett.\  {\bf B417}, 253 (1998)
[hep-th/9707241].

\bibitem{Russo:1998vt}
J.~G.~Russo,
Nucl.\ Phys.\  {\bf B535}, 116 (1998)
[hep-th/9802090].

\bibitem{GreenMoore} 
M.B.\ Green and G.\ Moore, unpublished.  

\bibitem{Green:1982sw}
M.~B.~Green, J.~H.~Schwarz and L.~Brink,
Nucl.\ Phys.\  {\bf B198}, 474 (1982).

\bibitem{Bel} 
L.\ Bel, Acad. Sci. Paris, Comptes Rend. {\bf 247}:1094 (1958), 
{\bf 248}:1297 (1959); I.\ Robinson, unpublished.  

\bibitem{D'Hoker:1995yr}
E.~D'Hoker and D.~H.~Phong,
Nucl.\ Phys.\  {\bf B440}, 24 (1995)
[hep-th/9410152].

\bibitem{Chalmers:1998dc}
G.~Chalmers,
Nucl.\ Phys.\  {\bf B524}, 295 (1998)
[hep-th/9712129].

\bibitem{D'Hoker:1988ta}
E.~D'Hoker and D.~H.~Phong,
Rev.\ Mod.\ Phys.\  {\bf 60}, 917 (1988).

\bibitem{Green:1999by}
M.~B.~Green and S.~Sethi,
Phys.\ Rev.\  {\bf D59}, 046006 (1999)
[hep-th/9808061].

\bibitem{Lechtenfeld:1990ke}
O.~Lechtenfeld and A.~Parkes,
Nucl.\ Phys.\  {\bf B332}, 39 (1990).

\bibitem{Harvey:1998ir}
J.~A.~Harvey and G.~Moore,
Phys.\ Rev.\  {\bf D57}, 2323 (1998)
[hep-th/9610237].

\bibitem{Gross:1986iv}
D.~J.~Gross and E.~Witten,
Nucl.\ Phys.\  {\bf B277}, 1 (1986).

\bibitem{Sakai:1987bi}
N.~Sakai and Y.~Tanii,
Nucl.\ Phys.\  {\bf B287}, 457 (1987).

\bibitem{Terras}  
A.\ Terras, Harmonic analyis on symmetric spaces and applications, I and II, 
Springer-Verlag, 1985.     

\bibitem{Howe:1984sr}
P.~S.~Howe and P.~C.~West,
Nucl.\ Phys.\  {\bf B238}, 181 (1984).

\bibitem{Cremmer:1981zg}
E.~Cremmer,
LPTENS 81/18 {\it Lectures given at ICTP Spring School Supergravity, Trieste, 
Italy, Apr 22 - May 6, 1981}.

\bibitem{Iengo:1999qh}
R.~Iengo and C.~Zhu,
JHEP {\bf 9906}, 011 (1999)
[hep-th/9905050].

\bibitem{Kiritsis:1997em}
E.~Kiritsis and B.~Pioline,
Nucl.\ Phys.\  {\bf B508}, 509 (1997)
[hep-th/9707018].

\bibitem{Obers:1999um}
N.~A.~Obers and B.~Pioline,
hep-th/9903113.

\bibitem{Kiritsis:2000zi}
E.~Kiritsis, N.~A.~Obers and B.~Pioline,
hep-th/0001083.

\bibitem{Bern:1997nh}
Z.~Bern, J.~S.~Rozowsky and B.~Yan,
Phys.\ Lett.\  {\bf B401}, 273 (1997)
[hep-ph/9702424].

\bibitem{Maldacena:1998re}
J.~Maldacena,
Adv.\ Theor.\ Math.\ Phys.\  {\bf 2}, 231 (1998)
[hep-th/9711200].

\bibitem{Gubser:1998bc}
S.~S.~Gubser, I.~R.~Klebanov and A.~M.~Polyakov,
Phys.\ Lett.\  {\bf B428}, 105 (1998)
[hep-th/9802109].

\bibitem{Witten:1998qj}
E.~Witten,
Adv.\ Theor.\ Math.\ Phys.\  {\bf 2}, 253 (1998)
[hep-th/9802150].



\end{thebibliography}
\end{document}